\documentclass[aps,prl,preprint,groupedaddress]{revtex4}
\usepackage{graphicx}
\usepackage{epsfig}
\usepackage{dcolumn}
\usepackage{bm}
\usepackage{subfigure}

\begin{document}

\title{Valley Kondo Effect in Silicon Quantum Dots }
\author{Shiue-yuan Shiau}
\author{Sucismita Chutia}
\author{Robert Joynt$^{\dagger }$}
\affiliation{Department of Physics, University of Wisconsin, 1150 Univ. Ave., Madison,
Wisconsin 53706, $\dagger $ and Department of Physics, University of Hong
Kong, Hong Kong, China}

\begin{abstract}
Recent progress in the fabrication of quantum dots using silicon opens the
prospect of observing the Kondo effect associated with the valley degree of
freedom. We compute the dot density of states using an Anderson model with
infinite Coulomb interaction $U$, whose structure mimics the nonlinear
conductance through a dot. The density of states is obtained as a function
of temperature and applied magnetic field in the Kondo regime using an
equation-of-motion approach. We show that there is a very complex peak
structure near the Fermi energy, with several signatures that distinguish
this spin-valley Kondo effect from the usual spin Kondo effect seen in GaAs
dots. We also show that the valley index is generally not conserved when
electrons tunnel into a silicon dot, though the extent of this
non-conservation is expected to be sample-dependent. We identify features of
the conductance that should enable experimenters to understand the interplay
of Zeeman splitting and valley splitting, as well as the dependence of
tunneling on the valley degree of freedom.
\end{abstract}

\maketitle

\section{I. Introduction}

Experimentation on gated quantum dots (QDs) has generally used GaAs as the
starting material, due to its relative ease of fabrication. However, the
spin properties of dots are becoming increasingly important, largely because
of possible applications to quantum information and quantum computing. Since
the spin relaxation times in GaAs are relatively short, Si dots, with much
longer relaxation times \cite{Jantsch2002, Tyryshkin2005,
Tyryshkin2003,DasSarma2003}, are of great scientific and technological
interest \cite{Eriksson2004}. Recent work has shown that few-electron
laterally gated dots can be made using Si \cite{Jones2006}\cite{Klein2004},
and single-electron QDs are certainly not far away.

There is one important qualitative difference in the energy level structures
of Si and GaAs: the conduction band minimum is two-fold degenerate in the
strained Si used for QDs as compared to the non-degenerate minimum in GaAs.
Thus each orbital level has a fourfold degeneracy including spin. This
additional multiplicity is referred to as the valley degeneracy. For
applications, this degeneracy poses challenges - it must be understood and
controlled. From a pure scientific viewpoint, it provides opportunities -
the breaking of the degeneracy is still poorly understood.

Previous experimental work has shown that in zero applied magnetic field the
splitting of the degeneracy is about $1.5\pm 0.6$ $\mu $eV in
two-dimensional electron gases (2DEGs) \cite{srijit_nature} and that it
increases when the electrons are further confined in the plane \cite%
{srijit_nature}. Surprisingly, the splitting increases linearly with applied
field. Theoretical understanding of these results has historically been
rather poor, with theoretical values much larger than experimental ones for
the zero-field splittings \cite{TBBoykin2004_2}\cite{TBBoykin2004} and
theory also predicting a nonlinear field dependence \cite{Ohkawa1977}.
Recent work indicates that consideration of surface roughness may resolve
these discrepancies \cite{Friesen2006}. In Si QDs, a related issue is also
of importance: what is the effect of valley degeneracy on the coupling of
the leads to the dots? The spin index is usually assumed to be conserved in
tunneling. Is the same true of the valley index?

\begin{figure}[t]
\epsfig{figure=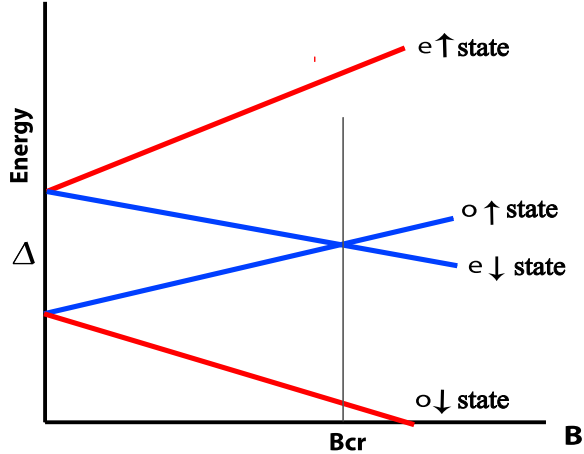,clip=,width=3 in}
\caption{Schematic diagram of the {energy levels of a single orbital in a Si
QD in a magnetic field. \ The quantum numbers are }{\protect\small {even~($e$%
), odd~($o$) valley states and up and down spin projections for spin $1/2$
states. A level crossing occurs at the magnetic field $B_{cr}$. }}}
\label{fig:fig1}
\end{figure}

Recent measurements of the valley splitting in a quantum point contact show
a valley splitting much larger than in the 2DEGs, about $1$ meV in \ quantum
point contacts \cite{srijit_nature}. A QD in a similar potential well is
expected to produce a valley splitting of the same order of magnitude. The
overall picture of the degeneracy is as shown in Fig.~\ref{fig:fig1} for a
single orbital level of a Si QD. Of particular interest is the fact that a
level crossing must occur, and the rough value of the applied field at this
point is $B_{cr}\approx 2.5$ T, given a reasonable zero-field valley
splitting of $0.5$ meV, and the valley-spin slope of magnetic field for the $%
"o\uparrow "$ state to be $0.1$ meV/T, as it is in Hall bars.

The aim of this paper is to study the Kondo effect in the context
of transport through a Si dot. The Kondo effect was originally
discovered in dilute magnetic alloys. At low temperatures, the
electron in a single impurity forms a spin singlet with electrons
in the conduction band, thus causing an increasing resistance as
the temperature is reduced to zero. Since the first observations
of the Kondo effect in GaAs QDs, there has been considerable
experimental and theoretical work done, mainly because QDs provide
an excellent playground where one can tune physical parameters
such as the difference between energy levels in QDs and the Fermi
level, the coupling to the leads, and the applied voltage
difference between the leads. The level of scientific
understanding of this purely spin Kondo effect in QDs is on the
whole quite satisfactory. The basic phenomena are as follows. At
temperature $T\leq T_{K}$, where the Kondo effect appears, a
zero-bias peak in the dot conductance is observed. An applied
magnetic field splits the peak into two peaks separated by twice
the Zeeman energy $2 g\mu _{B}B$. \ These linear peak energy
dependencies on the applied magnetic field have been observed in
GaAs QDs \cite{Goldhaber1998}. The Kondo effect only occurs when
the occupation of the dot is odd.
Clearly, Si dots will have a much richer phenomenology. There are multiple
field dependencies as seen in Fig.~\ref{fig:fig1}, and the additional
degeneracy can give rise to several Kondo peaks even in zero field.
Furthermore, the Kondo temperature is enhanced when the degeneracy is
increased. We build on previous work on the orbital Kondo effect. For
example, an enhanced Kondo effect has been observed due to extra orbital
degree of freedom in carbon nanotubes (CNTs) \cite{Jarillo}\cite{Choi2005}
and in a vertical QD, with magnetic field-induced orbital degeneracy for an
odd number of electrons and spin 1/2 \cite{Sasaki}. More complex peak
structure in their differential conductance suggests entangled interplay
between spin and orbital degrees of freedom. \ We also note that a zero-bias
peak has been observed in Si MOSFET structures \cite{tsui}.

Our goal will be to elucidate the characteristic structures in the
conductance, particularly their physical origin, and their temperature and
field dependencies. Once this is done, one can hope to use the Kondo effect
to understand some of the interesting physics of Si QDs, particularly the
dependence of tunneling matrix elements on the valley degree of freedom. \

This paper is organized as follows. In Sec.II, we determine whether valley
index conservation is to expected; in Sec. III, we introduce our formalism;
in Sec. IV, we discuss the effect of valley index non-conservation; in Sec.
V, we present our main results; in Sec. VI we summarize the implications for
experiments.

\section{II. Tunneling with valley degeneracy}

The basic issue at hand is the conductance of a dot with valley degeneracy,
and the influence of the Kondo effect on this process. \ The dots we have in
mind are separated electrostatically from 2DEG leads on either side. \ These
2DEGs have the same degeneracy structure. \ Thus before we tackle the issue
of the conductance we need the answer to a preliminary question: is the
valley index conserved during the tunneling process from leads to dots? \
Non-conservation of the valley index will introduce additional terms in the
many-body Hamiltonian that describes the dots, and, as we shall see below,
it changes the results for the conductance. \ In this section, we
investigate the question of valley index conservation in a microscopic model
of the leads and dots. \ We do this in the context of a single-particle
problem that should be sufficient to understand the parameters of the
many-body Hamiltonian that will be introduced in Sec. III.

If we have a system consisting of leads and a QD in a 2DEG of strained Si,
then both the leads and the dot will have even and odd valley states if the
interfaces defining the 2DEG are completely smooth. \ The valley state
degeneracy is split by a small energy. If the valley index is conserved
during tunneling, then the even valley state in the lead will only tunnel
into the even valley state in the dot. On the other hand, if the valley
index is not conserved, then the even valley state in the lead will tunnel
into both the even and odd valley states in the dot; similarly for the odd
valley state in the lead. We can estimate the strength of the couplings
between the various states in the leads and the dot by calculating the
hopping matrix elements between them.

Consider a system of a lead and a dot in a SiGe/Si/SiGe quantum well,
separated by a barrier of height $V_{b}$. If $V_{b}=\infty $, there will be
no tunneling between the lead and the dot and the eigenstates of the whole
system can be divided into lead states $|\Psi _{L,(e,o)}^{\infty }\rangle $
and dot states $|\Psi _{D,(e,o)}^{\infty }\rangle $ where $e,~o$ are the
even and the odd valley states. Call the Hamiltonian for this case $H_{0}$.
When $V_{b}$ is lowered to a finite value, there will be some amount of
tunneling of the lead wavefunction into the dot. Call the Hamiltonian for
this case $H$. This can be thought of as a perturbation problem where the $%
n^{th}$ perturbed eigenstate $|n\rangle $ is no longer the $n^{th}$
unperturbed eigenstate $|n^{0}\rangle $ but acquires components along the
other unperturbed eigenstates $|k^{0}\rangle $. The perturbation
Hamiltonian is $H^{\prime }=H-H_{0}.$ \ $H^{\prime }$ considered as a
function of position is not small, but the matrix elements of \thinspace $%
H^{\prime }$ are small. This means that in the $1^{st}$ order perturbation
theory we can write the perturbed $n^{th}$ state as
\begin{equation}
|n\rangle =|n^{0}\rangle +\sum_{k\neq n}|k^{0}\rangle \frac{V_{kn}}{%
E_{n}^{0}-E_{k}^{0}}
\end{equation}%
where $V_{kn}=\langle k^{0}|H^{\prime }|n^{0}\rangle $ is the hopping or
tunneling matrix term.

If we further assume that the tunneling Hamiltonain conserves spin and that
only one orbital state need be considered on the QD, we can expand the
perturbed even valley lead wavefunction (with tunneling) as
\begin{equation}
|\Psi _{L,e}\rangle =|\Psi _{L,e}^{\infty }\rangle +\frac{V_{e,e}}{%
|E_{Le}^{0}-E_{De}^{0}|}|\Psi _{D,e}^{\infty }\rangle +\frac{V_{e,o}}{%
|E_{Le}^{0}-E_{Do}^{0}|}|\Psi _{D,o}^{\infty }\rangle
\end{equation}%
Usually, in a perturbation problem, the unperturbed wavefunctions and the
hopping matrix elements are known and used to find the perturbed
wavefunction. However in the present case we use a model to compute the
perturbed and unpertubed wavefunctions and use these to determine the
unknown hopping matrix elements $V_{e,e}$ $V_{e,o}.$ \ These will be needed in
later sections.

From the above equation,
\begin{eqnarray}
V_{e,e}&=&(|E_{Le}^0-E_{De}^0|)\langle \Psi^\infty_{De}|\Psi_{Le}\rangle \\
V_{e,o}&=&(|E_{Le}^0-E_{Do}^0|)\langle \Psi^\infty_{Do}|\Psi_{Le}\rangle
\end{eqnarray}
The perturbed and the unperturbed wavefunctions and the corresponding
energies are calculated by using an empirical 2D tight binding model with
nearest ($v_{z}$) and next nearest neighbor interactions ($u_{z}$) along the
$z$ (growth) direction and the nearest neighbor interactions ($v_{x}$) along
the $x$ direction (in the plane of the 2DEG). Potential barriers and edges
of the quantum well are modeled by adjusting the onsite parameter ($\epsilon
$) on the atoms. This is an extension of the 2-band 1D tight binding model
outlined by Boykin\textit{\ et al.} \cite{TBBoykin2004_2,TBBoykin2004},
considering only the lowest conduction band of Si. This is a single particle
calculation with the magnetic field $B=0$.\ Note that the parameters from the
Boykin\textit{\ et al}. model are chosen precisely so as to reproduce the
two-valley structure of strained Si.\

\begin{figure}[h]
\begin{center}
\includegraphics[width=4.5in]{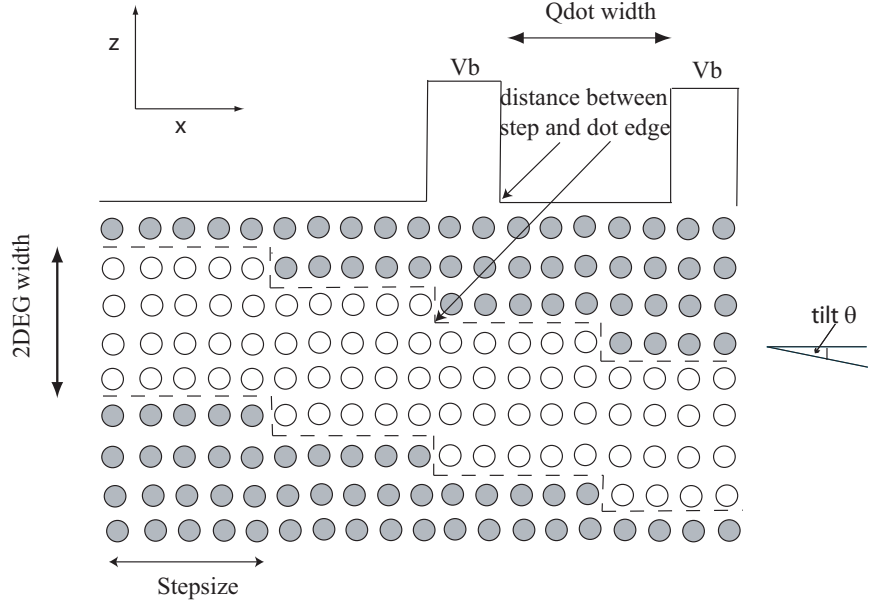}
\end{center}
\caption{The system of the lead and the dot in a miscut quantum well with
the potential profile along the $x$-direction. The open circles represent Si
atoms sandwiched between the SiGe atoms in the barrier region represented by
the filled circles.}
\label{MiscutWell}
\end{figure}

The single-particle Hamiltonian for the system is
\begin{eqnarray}
H &=&\sum_{x}~[~\epsilon |\phi (x,z)\rangle \langle \phi (x,z)|+v_{x}|\phi
(x,z)\rangle \langle \phi (x+1,z)|+v_{z}|\phi (x,z)\rangle \langle \phi
(x,z+1)|  \nonumber \\
&+&u_{z}|\phi (x,z)\rangle \langle \phi (x,z+2)|+v_{x}|\phi (x-1,z)\rangle
\langle \phi (x,z)|+v_{z}|\phi (x,z-1)\rangle \langle \phi (x,z)|  \nonumber
\\
&+&u_{z}|\phi (x,z-2)\rangle \langle \phi (x,z)|~]  \label{eq:2}
\end{eqnarray}%
The parameters defining the system are as follows: the lead is $80$ atoms
long, the barrier between the lead and the dot is $16$ atoms wide along $x$
and $0.3$ eV high, the QD is $38$ atoms ($\approx 10~$nm) wide along $x$ and
$38$ atoms wide along $z$, the barrier along $z$ defining the quantum well
is $20$ atoms wide on each side and its height is $0.3$ eV. The nearest and
next nearest neighbor interaction terms along $z$ are $v_{z}=0.68264$ eV,
and $u_{z}=0.611705$ eV, and that along $x$ is $v_{x}=-10.91$ eV. \ The
schematic of the system is shown in Fig.~\ref{MiscutWell}

\begin{figure}[h]
\begin{center}
\includegraphics[width=4.5in]{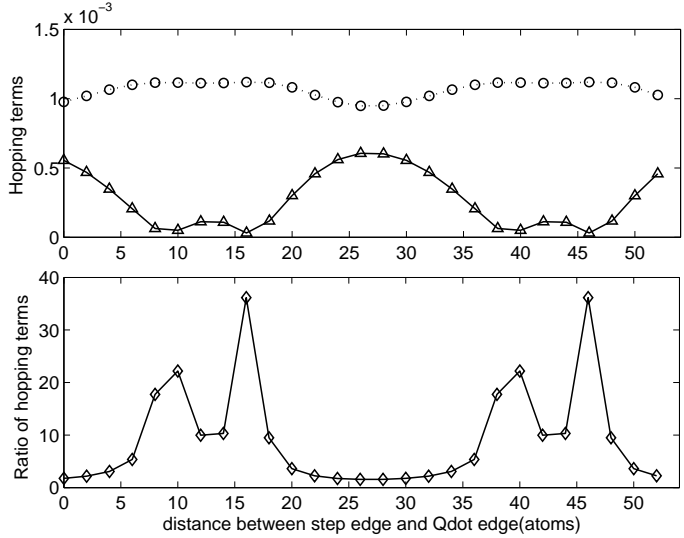}
\end{center}
\caption{(a)Tunneling matrix elements ($V_{e,e}, V_{e,o}$) in units of eV
and (b) Ratio of the tunneling matrix elements ($V_{e,o}/V_{e,e}$) as a
function of the relative distance between the step edge and the quantum dot
edge for constant barrier height and width (for a miscut QW). The solid line
with triangular markers indicate the $V_{e,e}$ term and the dashed line with
open circles indicate the $V_{e,o}$ term. Here the stepsize is constant at $%
30$ atoms ($\approx 2^o$ tilt)}
\label{TMvsDis}
\end{figure}

\begin{figure}[h]
\begin{center}
\includegraphics[width=4.5in]{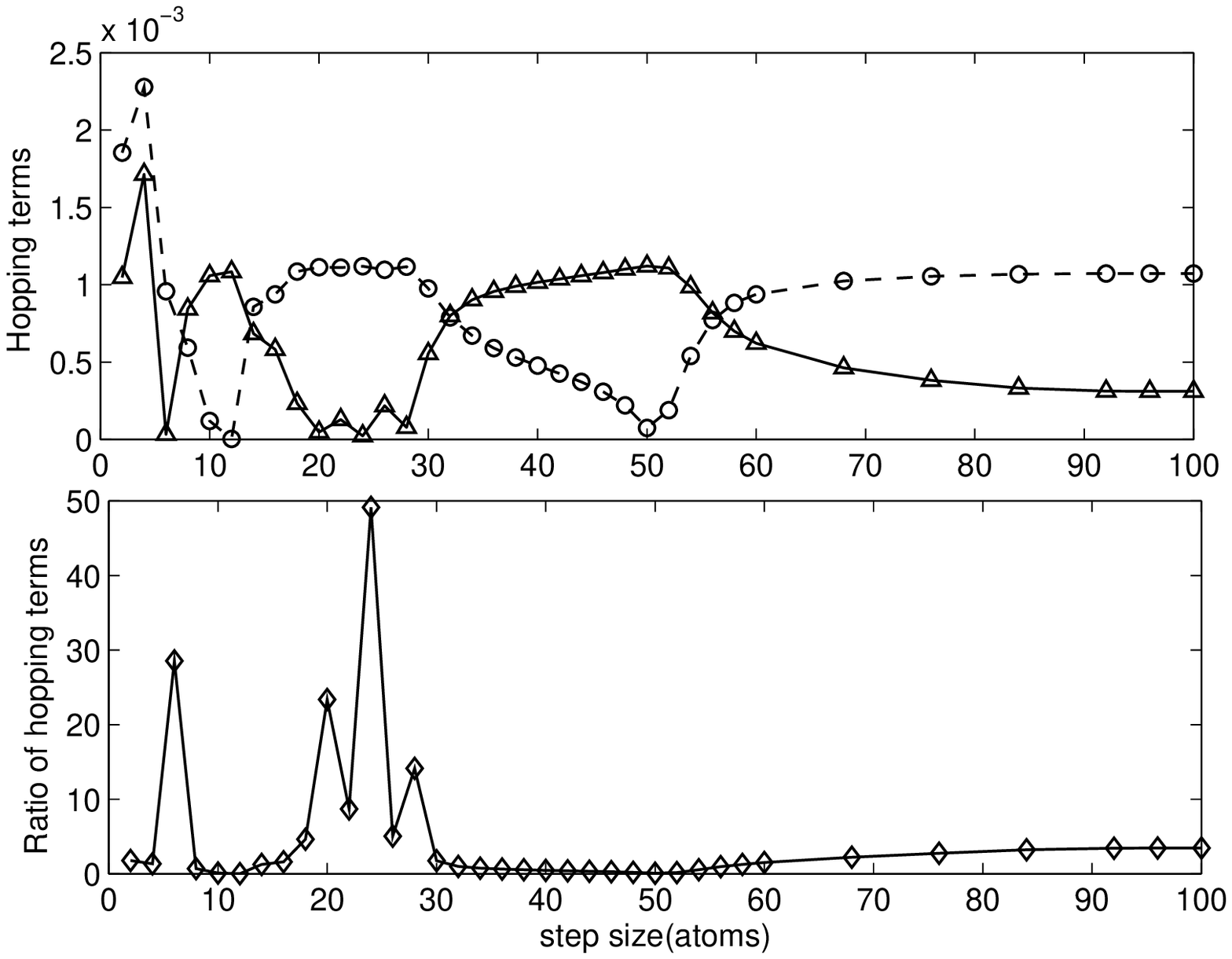}
\end{center}
\caption{(a)Tunneling matrix elements ($V_{e,e}, V_{e,o}$) in units of eV
and (b) Ratio of the tunneling matrix elements ($V_{e,o}/V_{e,e}$) as a
function of the stepsize for constant barrier height and width (for a miscut
QW). The solid line with triangular markers indicate the $V_{e,e}$ term and
the dashed line with open circles indicate the $V_{e,o}$ term. Here the step
edge coincides with the quantum dot edge. }
\label{TMvsSs}
\end{figure}

We consider two kinds of 2DEGs, one with no interface roughness at the
interface and the other with a miscut of about $2^{\circ}$. We find that the
valley index is conserved in the case of smooth interfaces. We model the
miscut interface as a series of regular steps of one atom thickness and
width defined by the tilt, $e.g.,$ a $2^{\circ }$ tilt corresponds to a step
length of about $30$ Si atoms. In this case, the valley index is no longer
conserved. Strengths of the coupling between the $e-e$ and $e-o$ valley
states is found to depend strongly on the relative distance between the edge
of the closest step and the edge of the quantum dot. In Fig.~\ref{TMvsDis},
we plot the hopping matrix elements as a function of this relative distance.
The terms show oscillations that have a period equal to the step size in
this case.

The coupling strengths are also found to depend on the tilt of the substrate
on which the 2DEG is grown. We show in Fig.~\ref{TMvsSs}, the hopping matrix
elements as a function of the stepsize ($\propto $ 1/ tilt) and find that
the hopping terms vary rapidly at small step sizes ($i.e$ large tilts) but
slowly at larger stepsizes ($i.e.,$ smaller tilts).

In a quantum well, the phases of the rapid oscillations of the electron wavefunctions
(along the z-direction) are locked to the interface and in a well with
smooth interfaces, the phase remains the same for a particular valley
throughout \cite{friesen1}. But in a miscut well, the phases are different
for electrons localized at different steps even for the same valley
eigenstate. The resultant valley splitting is due to interference of phase
contributions from all the steps. Hence the valley splitting can be very
different in the lead and the dot. The same argument explains the
conservation of the valley index across the barrier in the lead-dot system.
In the case of a smooth interface, the $
e$ eigenstate in the lead has same phase of the
z-oscillations as the $ e$ eigenstate in
the dot, and is $90^{\circ }$ out of phase with the $
o$ eigenstate in the dot. Hence, there is a finite overlap
between the $ e-e $ wavefunctions, while
the $e-o$ overlap cancels out due to interference. This preserves the valley
index during the tunneling.

In the case of a miscut well, we can no longer strictly label the
eigenstates $e$ or $o$, but let us continue to do so just for
convenience. Here, because $e$ in the lead does not really have a
single phase along $z$ but is composed of different phases at each
step and similarly for $o$ in the dot, there is a fair chance that
there will be a non-cancellation of the phases between the $e,~o$
wavefunctions depending up the the extent of tunneling. This will
give rise to a finite hopping matrix term between all the dot and
lead states and the valley index $o,~e$ can longer be said to be
conserved.

The results show that the conservation or non-conservation of the valley
index depends sensitively on fine details of the dot morphology such the
proximity of a step edge to the edge of the dot on atomic scales. \ It is
not likely that good control at this level can be achieved, and thus some
sample dependence must be expected. \ This means that it will normally be
necessary to include hopping terms that do not conserve valley index into
the Hamiltonian in order to understand the Kondo effect in realistic Si QDs,
where the interfaces are rarely smooth.

\section{III. Equation-of-Motion Approach in the $U\rightarrow \infty $ limit%
}

Sec. II has given us some insight into the nature of the tunneling between
the leads and the dot. \ We now introduce a Hamiltonian to describe the full
many-body problem, and discuss the computational method we will use to solve
it. \ This Hamiltonian must include the single-particle energy levels of the
leads and the dot, the tunneling matrix elements that connect these levels,
and the Coulomb interaction between electrons on the dot. \ We shall use the
Anderson impurity model:
\begin{eqnarray}
\mathcal{H} &=&\sum_{ikm\sigma }\varepsilon _{k}c_{ikm\sigma
}^{+}c_{ikm\sigma }+\sum_{m\sigma }\varepsilon _{m\sigma }f_{m\sigma
}^{+}f_{m\sigma }+\sum_{ikm\sigma }V_{0,ik}(c_{ikm\sigma }^{+}f_{m\sigma
}+f_{m\sigma }^{+}c_{ikm\sigma })  \nonumber \\
&&+\sum_{ikm\sigma }V_{X,ik}(c_{ikm\sigma }^{+}f_{\bar{m}\sigma }+f_{\bar{m}%
\sigma }^{+}c_{ikm\sigma })+\frac{U}{2}\sum_{m^{\prime }\sigma ^{\prime
}\not=m\sigma }n_{m^{\prime }\sigma ^{\prime }}n_{m\sigma }  \nonumber
\end{eqnarray}%
Here $\sigma $ indicate two-fold spin $1/2$ and $m$ two-fold valley indices.
$\bar{m}$ means the opposite valley state of $m$. The operator $c_{ikm\sigma
}^{+}(c_{ikm\sigma })$ creates (annihilates) an electron with an energy $%
\varepsilon _{k}$ in the $i$ lead, $i\in L,R$, while the operator $%
f_{m\sigma }^{+}(f_{m\sigma })$ creates (annihilates) an electron with an
energy $\varepsilon _{m\sigma }$ in the QD, connected to the leads by
Hamiltonian couplings $V_{0,ik}$ and $V_{X,ik}$ which correspond to an
electron tunneling between the same valley states and different valley
states, respectively. We assume that the $V_{0(X),ik}$ do not depend on spin
index $\sigma $. $U$ is the Coulomb interaction on the dot. \ It is assumed
to be independent of $m.$ \ Previous studies on the orbital Kondo effect
have mostly assumed conservation of orbital indices \cite{orbitkondo}. \
Recent theoretical work has found that a transition occurs from $SU(4)$ to $%
SU(2)$ Kondo effect by allowing violation of conservation of orbital index
in CNTs \cite{Choi2005,JSLim2006} or in double QDs \cite{Chudnovskiy2005}.
For Si QDs, we have shown from the previous section that tunnelings between
different valley states have to be taken into account.

Experiments measure the current $I$ as a function of source-drain voltage $%
V_{sd}.$ \ The differential conductance $G=d\mathrm{I}/d\mathrm{V}_{sd}$ is
given by the generalized Landauer formula \cite{Landauer}
\[
\mathrm{I}=\frac{e}{h}\sum_{m\sigma }\int_{-D}^{D}dw(f_{L}(w)-f_{R}(w))\frac{%
2\Gamma _{L,m\sigma }(w)\Gamma _{R,m\sigma }(w)}{\Gamma _{L,m\sigma
}(w)+\Gamma _{R,m\sigma }(w)}\mathrm{Im}[\mathcal{G}_{m\sigma }(w)]
\]%
where $D$ is the bandwidth and $\Gamma _{i,m\sigma }(w)=2\pi
\sum_{k}(V_{0,ik}^{2}+V_{X,ik}^{2})\delta (w-\varepsilon _{k})$. The $%
f^{\prime }s$ are Fermi functions calculated with chemical potentials $\mu
_{L}$ and $\mu _{R}$ with $\mu _{L}=\mu _{R}+e\mathrm{V}_{sd}.$ \ In the
following argument, we assume a flat unperturbed density of states and $%
V_{0,ik}=V_{0},~V_{X,ik}=V_{X}$. We then define $\Gamma =\Gamma _{L,m\sigma
}(w)+\Gamma _{R,m\sigma }(w)=\pi V^{2}/D$ with $V^{2}=V_{0}^{2}+V_{X}^{2}$.
\ \ With these approximations we may differentiate this equation and we find
$d\mathrm{I}/d\mathrm{V}_{sd}\sim $~$-\mathrm{Im}[\mathcal{G}_{m\sigma }(e%
\mathrm{V}_{sd})]/\pi =$ density of states(DOS). \ Here $\mathcal{G}%
_{m\sigma }(w)$ is the retarded Green's function:
\begin{equation}
\mathcal{G}_{m\sigma }(w)=\ll f_{m\sigma },f_{m\sigma }^{+}\gg
=-i\int_{0}^{\infty }e^{i\alpha t}<\{f_{m\sigma }(t),f_{m\sigma
}^{+}(0)\}>dt.  \label{eq:f0}
\end{equation}%
where $\alpha =w+i\delta $. Here $\{,\}$ and $<,>$ denote the
anti-commutator and statistical average of operators, respectively. \ Hence
our task is to compute $\mathcal{G}_{m\sigma }(w).$ \ Although the
approximation of a flat density of states is not likely to be completely
valid, we still expect that sharp structures in $\mathcal{G}_{m\sigma }(w)$
will be reflected in the voltage dependence of $d\mathrm{I}/d\mathrm{V}_{sd}%
\mathrm{.}$

Several approximate solution methods for this type of Hamiltonian have been
used successfully: numerical renormalization group, non-crossing
approximation(NCA), scaling theory, and equation-of-motion (EOM) approach.
In this paper, we use the EOM approach to investigate the transport through
a Si QD at low temperature, in the presence of magnetic field. The EOM
approach has several merits. The most important for our purposes is that it
can produce the Green's function at finite temperature, and works both for
infinite-$U$ and finite-$U$. We basically follow the spirit of the paper by
Czycholl \cite{Czycholl} which gives a thorough analysis of EOM in the large-%
$N$ limit($N$ is the number of energy level degeneracy) and provides a
suitable approximation for finite-$U$ systems. Though in the present case $N$
is $4$, the Green's function obtained from EOM should give a reasonable
picture in the case of a weak hybridization and relatively large $U$. \ The
approximation starts from the equations of motion for $\mathcal{G}_{m\sigma
}(w)$ in the frequency domain:
\[
w\ll A,B\gg =<\{A,B\}>+\ll \lbrack A,H],B\gg =<\{A,B\}>+\ll A,[H,B]\gg .%
\nonumber
\]%
This can be obtained by integrating Eq.~\ref{eq:f0} by parts. The
detailed derivation for the Green's function $\mathcal{G}_{m\sigma
}$ is given in the Appendix. We truncate higher-order Green's
functions by using a scheme adopted by
\cite{Lacroix1981,Luo2002,Kashcheyevs2006}. The higher-order
Green's functions are of the general forms $\ll f^{+}c c,f^{+}\gg
$ and $\ll fc^{+}c,f^{+}\gg $ (we drop indices for the sake of
generalization) which we decouple by replacing them with
\begin{eqnarray}
\ll f^{+}c c,f^{+}\gg  &=&<f^{+}c>\ll c,f^{+}\gg   \nonumber \\
\ll fc^{+}c,f^{+}\gg  &=&<c^{+}c>\ll f,f^{+}\gg .  \nonumber
\end{eqnarray}%
We combine Eq.~\ref{eq:aa1},~\ref{eq:aa2},~\ref{eq:rf4}, and \ref{eq:rff4}
and take the limit $U\rightarrow \infty $. We obtain coupled Green's
functions expressed by
\begin{eqnarray}
R(z)\mathcal{G}_{m\sigma } &=&1-\sum_{l\not=m\sigma
}<n_{l}>-\sum_{l\not=m\sigma }\tilde{A}_{l}(z)+P(z)\mathcal{M}_{\bar{m}%
\sigma }  \nonumber \\
\bar{R}(\bar{z})\mathcal{M}_{\bar{m}\sigma } &=&<f_{m\sigma }^{+}f_{\bar{m}%
\sigma }>+\widetilde{\bar{F}}_{m\sigma }(w)+\bar{P}(\bar{z})\mathcal{G}%
_{m\sigma },  \label{eq:fbarf}
\end{eqnarray}%
where $\mathcal{M}_{\bar{m}\sigma }\equiv \ll f_{\bar{m}\sigma },f_{m\sigma
}^{+}\gg $ is the correlation function between different valley states. We
denote $z=w-\varepsilon _{m\sigma }+\varepsilon _{l\not=m\sigma }$ and $\bar{%
z}=w-\varepsilon _{\bar{m}\sigma }+\varepsilon _{p\not=\bar{m}\sigma }$. The
propagators $R(z),~\bar{R}(\bar{z})$ and $~P(z)$, $\bar{P}(\bar{z})$ are
defined as follows
\begin{eqnarray}
R(z) &=&w-\varepsilon _{m\sigma }-\Sigma _{a}(w)(1-\sum_{l\not=m\sigma }%
\tilde{A}_{l}(z))-\sum_{l\not=m\sigma }\Sigma _{c,l}(z)-\Sigma _{b}(w)%
\widetilde{F}_{m\sigma }(w)-\sum_{l\not=m\sigma }\widetilde{J}_{l}(z)
\nonumber \\
\bar{R}(\bar{z}) &=&w-\varepsilon _{\bar{m}\sigma }-\Sigma
_{a}(w)(1-\sum_{p\not=\bar{m}\sigma }\tilde{A}_{p}(\bar{z}))-\sum_{p\not=%
\bar{m}\sigma }\Sigma _{c,p}(\bar{z})-\Sigma _{b}(w)\widetilde{\bar{F}}%
_{m\sigma }(w)-\sum_{p\not=\bar{m}\sigma }\widetilde{J}_{p}(\bar{z})
\nonumber \\
P(z) &=&\Sigma _{b}(w)(1-\sum_{l\not=m\sigma }\tilde{A}_{l}(z))-\Sigma
_{d}(w)-\tilde{I}_{m\sigma }(w)+\Sigma _{a}(w)\widetilde{F}_{m\sigma }(w)
\nonumber \\
\bar{P}(\bar{z}) &=&\Sigma _{b}(w)(1-\sum_{p\not=\bar{m}\sigma }\tilde{A}%
_{p}(\bar{z}))-\Sigma _{d}(w)-\tilde{\bar{I}}_{m\sigma }(w)+\Sigma _{a}(w)%
\widetilde{\bar{F}}_{m\sigma }(w).  \nonumber
\end{eqnarray}%
Note that $l\not=m\sigma ,~p\not=\bar{m}\sigma $. The various functions in $%
R(z),~\bar{R}(\bar{z}),~P(z)$, and $\bar{P}(\bar{z})$ are defined in Eq.~\ref%
{eq:deffunction_1}-\ref{eq:deffunction_10} in the Appendix. Moreover, they
can be self-consistently computed. We proceed to solve for $\mathcal{G}%
_{m\sigma }$ and $\mathcal{M}_{\bar{m}\sigma }$
\begin{eqnarray}
\mathcal{G}_{m\sigma } &=&\frac{1-\sum_{l\not=m\sigma
}<n_{l}>-\sum_{l\not=m\sigma }\tilde{A}_{l}(z)+(<f_{m\sigma }^{+}f_{\bar{m}%
\sigma }>+\widetilde{\bar{F}}_{m\sigma }(w))P(z)/\bar{R}(\bar{z})}{R(z)-P(z)%
\bar{P}(\bar{z})/\bar{R}(\bar{z})}  \label{eq:gref1} \\
\mathcal{M}_{\bar{m}\sigma } &=&\frac{<f_{m\sigma }^{+}f_{\bar{m}\sigma }>+%
\widetilde{\bar{F}}_{m\sigma }(w)+(l-\sum_{l\not=m\sigma
}<n_{l}>-\sum_{l\not=m\sigma }\tilde{A}_{l}(z))\bar{P}(\bar{z})/R(z)}{\bar{R}%
(\bar{z})-\bar{P}(\bar{z})P(z)/R(z)}.  \label{eq:gref2}
\end{eqnarray}%
$P(z),~\bar{P}(\bar{z})$ express the correlations between $"m\sigma "$ and $"%
\bar{m}\sigma "$ states. If $V_{X}=0$, then $P(z)=0$, Coupled Eqs.~\ref%
{eq:fbarf} are decoupled and reduced to the Green's function with conserved
valley index. Furthermore, when $V_{X}\not=0$, peak structure will alter
since the function $\Sigma _{d}(w)$ in $P(z),~\bar{P}(\bar{z})$ has a
second-order logarithmic divergence at the Fermi level. %\begin{eqnarray}
%G_m(w)=\frac{1-\sum_{l\not= m}(<n_l>+A_l(w+\varepsilon_l-\varepsilon_m ))}{w-\varepsilon_m-\Sigma_0(w)-\sum_{l\not= m}\Sigma_{1,ml}(w)}\label{eq:gf1}
%\end{eqnarray}
%where we define
%\begin{eqnarray}
%\Sigma_0(w) &=&\sum_{ik} \frac{V_{ik}^2}{w-\varepsilon_k}\nonumber\\
% \Sigma_{1,ml}(w) &=&\sum_{ik} \frac{V_{ik}^2 f_{FD}(\varepsilon_k )}{w-\varepsilon_k
%+\varepsilon_l -\varepsilon_m }\nonumber
%\end{eqnarray}
Now we have a\ set of self-consistent equations for $\mathcal{G}_{m\sigma }$
from Eq.~\ref{eq:gref1}, since $\tilde{A}_{m\sigma }(z)$, $<n_{m\sigma
}>,~<f_{m\sigma }^{+}f_{\bar{m}\sigma }>$ and other terms such as $\tilde{I}%
_{m\sigma }(w),~\tilde{F}_{m\sigma }(w),~\tilde{J}_{m\sigma }(z)$ are
integrals over $\mathcal{G}_{m\sigma }$ and $\mathcal{M}_{\bar{m}\sigma }$
themselves. We solve by taking an initial guess for $\mathcal{G}_{m\sigma }$
and $\mathcal{M}_{\bar{m}\sigma }$ of the forms
\begin{eqnarray}
\mathcal{G}_{m\sigma }^{0}(w) &=&\frac{1-\sum_{l\not=m\sigma }<n_{l}>}{%
w-\varepsilon _{m\sigma }+iN\Gamma \pi }.  \nonumber \\
\mathcal{M}_{\bar{m}\sigma }^{0}(w) &=&\frac{<f_{m\sigma }^{+}f_{\bar{m}%
\sigma }>}{w-\varepsilon _{\bar{m}\sigma }+iN\Gamma \pi }.  \nonumber
\end{eqnarray}%
These have the expected structure of a broad peak of width $N\Gamma $ $(N=4)$
centered around the energy $\varepsilon _{m\sigma }$ for $\mathcal{G}%
_{m\sigma }$ and $\varepsilon _{\bar{m}\sigma }$ for $\mathcal{M}_{\bar{m}%
\sigma }$, and the narrow Kondo peak(s) around the Fermi level. \ \ We then
iterate to self-consistency, at each stage determining the occupation
numbers $<n_{m\sigma }>$ and expectation values $<f_{m\sigma }^{+}f_{\bar{m}%
\sigma }>$ from\cite{foot_note1}
\begin{eqnarray}
<n_{m\sigma }> &=&-\frac{1}{\pi }\int dwf_{FD}(w)\mathrm{Im}\mathcal{G}%
_{m\sigma }.  \nonumber \\
<f_{m\sigma }^{+}f_{\bar{m}\sigma }> &=&-\frac{1}{\pi }\int dwf_{FD}(w)%
\mathrm{Im}\mathcal{M}_{\bar{m}\sigma }.  \nonumber
\end{eqnarray}%
Here $f_{FD}$ is the Fermi function.

\section{IV. Effects of non-conservation of valley index}

Consider all four-fold valley and spin levels to be degenerate, meaning no
magnetic field and no zero-field valley splitting. We have $R(z)=\bar{R}(%
\bar{z})$, $P(z)=\bar{P}(\bar{z})$ and $z=\bar{z}=w$. We define the Kondo
temperature as the temperature at which the real part of the denominator of
the Green's function vanishes. As discussed by Hewson \cite{Hewson}, this
definition gives the temperature that controls the thermodynamics. \ We
rewrite Eq.~\ref{eq:gref1} in order to explicitly express two contributions
from $\mathcal{G}_{m\sigma }$,
\begin{eqnarray}
\mathcal{G}_{m\sigma } &=&\frac{1-\sum_{l\not=m\sigma
}<n_{l}>-\sum_{l\not=m\sigma }\tilde{A}_{l}(w)+<f_{m\sigma }^{+}f_{\bar{m}%
\sigma }>+\widetilde{\bar{F}}_{m\sigma }(w)}{2(R(w)-P(w))}  \nonumber \\
&&+\frac{1-\sum_{l\not=m\sigma }<n_{l}>-\sum_{l\not=m\sigma }\tilde{A}%
_{l}(w)-(<f_{m\sigma }^{+}f_{\bar{m}\sigma }>+\widetilde{\bar{F}}_{m\sigma
}(w))}{2(R(w)+P(w))}.  \label{eq:gref1_1}
\end{eqnarray}%
We hence find two Kondo temperatures corresponding to the solutions of
\begin{eqnarray}
\mathrm{Re}R(w)+\mathrm{Re}P(w) &=&0,\text{and}  \nonumber \\
\mathrm{Re}R(w)-\mathrm{Re}P(w) &=&0.  \nonumber
\end{eqnarray}%
To second order in the $V$'s, these equations are
\begin{eqnarray}
w-\varepsilon _{m\sigma }-\sum_{ik}f_{FD}(\varepsilon _{k})\frac{%
2V_{0}^{2}+2V_{X}^{2}+(V_{0}+V_{X})^{2}}{w-\varepsilon _{k}} &=&0~\text{and}
\nonumber \\
w-\varepsilon _{m\sigma }-\sum_{ik}f_{FD}(\varepsilon _{k})\frac{%
2V_{0}^{2}+2V_{X}^{2}+(V_{0}-V_{X})^{2}}{w-\varepsilon _{k}} &=&0.  \nonumber
\end{eqnarray}%
In general each equation has three solutions, but only one of them lies
above the cut $-D<w<0$ on the real axis. To find $T_{K}$, we set $%
w=k_{B}T_{K}$, $\varepsilon _{F}=0$ and assume $k_{B}T_{K}\ll D$. As the temperature tends to zero, $T_K$'s satisfy
\begin{eqnarray}
k_{B}T_{K1} &=&\varepsilon _{m\sigma }-\frac{\Gamma _{1}}{\pi }\mathrm{ln}%
\left\vert \frac{k_{B}T_{K1}}{D}\right\vert   \nonumber \\
k_{B}T_{K2} &=&\varepsilon _{m\sigma }-\frac{\Gamma _{2}}{\pi }\mathrm{ln}%
\left\vert \frac{k_{B}T_{K2}}{D}\right\vert ,  \nonumber
\end{eqnarray}%
where
\begin{eqnarray}
\Gamma _{1} &=&\frac{\pi }{D}\left[ 2V_{0}^{2}+2V_{X}^{2}+(V_{0}+V_{X})^{2}%
\right]   \nonumber \\
\Gamma _{2} &=&\frac{\pi }{D}\left[ 2V_{0}^{2}+2V_{X}^{2}+(V_{0}-V_{X})^{2}%
\right] .  \nonumber
\end{eqnarray}%
Define $V_{0}=V\cos {\phi },~V_{X}=V\sin {\phi },~\beta =\sin {2\phi }$,
with $0\leq \beta \leq 1$. \ $\beta =0$ implies valley index conservation,
while $\beta =1$ means that tunnelings between same valley states and
different valley states both take place.

We express $\Gamma _{1},~\Gamma _{2}$ as
\begin{eqnarray}
\Gamma _{1} &=&\frac{\pi V^{2}}{D}(3+\beta )=(3+\beta )\Gamma  \nonumber \\
\Gamma _{2} &=&\frac{\pi V^{2}}{D}(3-\beta )=(3-\beta )\Gamma  \nonumber
\end{eqnarray}%
with $\Gamma =\pi V^{2}/D$. Assuming $k_{B}T_{K1},k_{B}T_{K2} \ll
\varepsilon _{m\sigma } $, we obtain
\begin{eqnarray}
k_{B}T_{K1} &=&D\exp {(\pi \varepsilon _{m\sigma }/(3+\beta )\Gamma )}
\label{eq:kondoTK1} \\
k_{B}T_{K2} &=&D\exp {(\pi \varepsilon _{m\sigma }/(3-\beta )\Gamma )}
\label{eq:kondoTK2}
\end{eqnarray}

There are two important observations from the above result: first, the Kondo
temperature depends exponentially on $\beta $; secondly, there are two Kondo
temperatures which split when $\beta >0$. For the case $\beta =0$
(conservation of valley index) we find $T_{K1}=T_{K2}=D\exp {(\pi
\varepsilon _{m\sigma }/3\Gamma )}$, as expected from the EOM approach. When
$0<\beta <1$, meaning either $V_{0}>V_{X}$ or $V_{0}<V_{X}$, electrons can
now tunnel through same valley states or different valley states and the two
Kondo temperatures $T_{K1}(\beta )$ and $T_{K2}(\beta )$ split. When $\beta
=1$, we have the maximum $T_{K1}(\beta =1)=(DT_{K2}(\beta
=1))^{1/2}>T_{K}(\beta =0)>T_{K2}(\beta =1)$.

\begin{figure}[t]
\vspace{-3 cm} \centering
\subfigure[~$\beta=0$]  {\
\includegraphics[width=.34\textwidth]{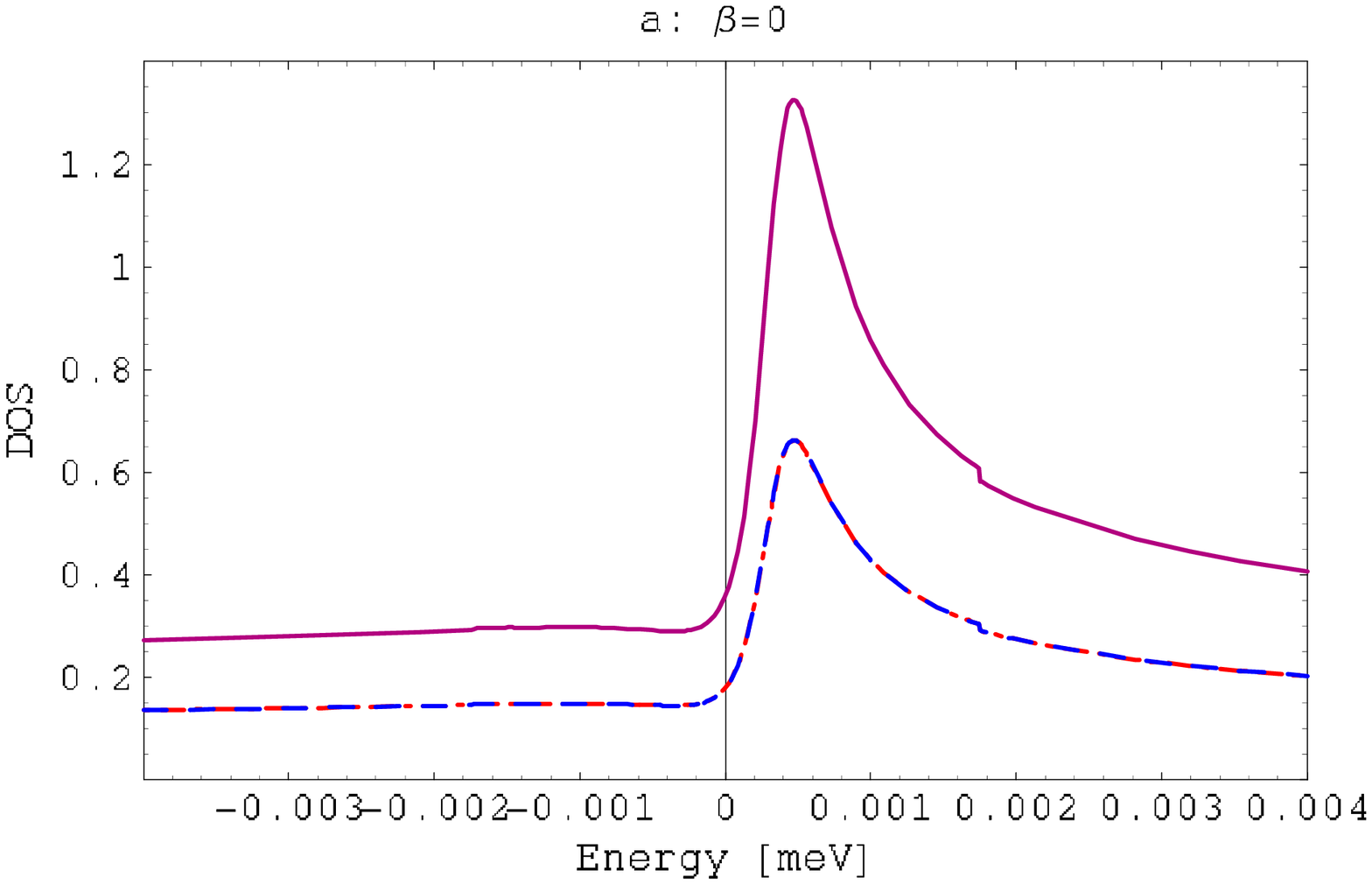}} \hspace{-0.7 cm}
\subfigure[~$\beta=0.5$]  {\
\includegraphics[width=.33\textwidth]
                {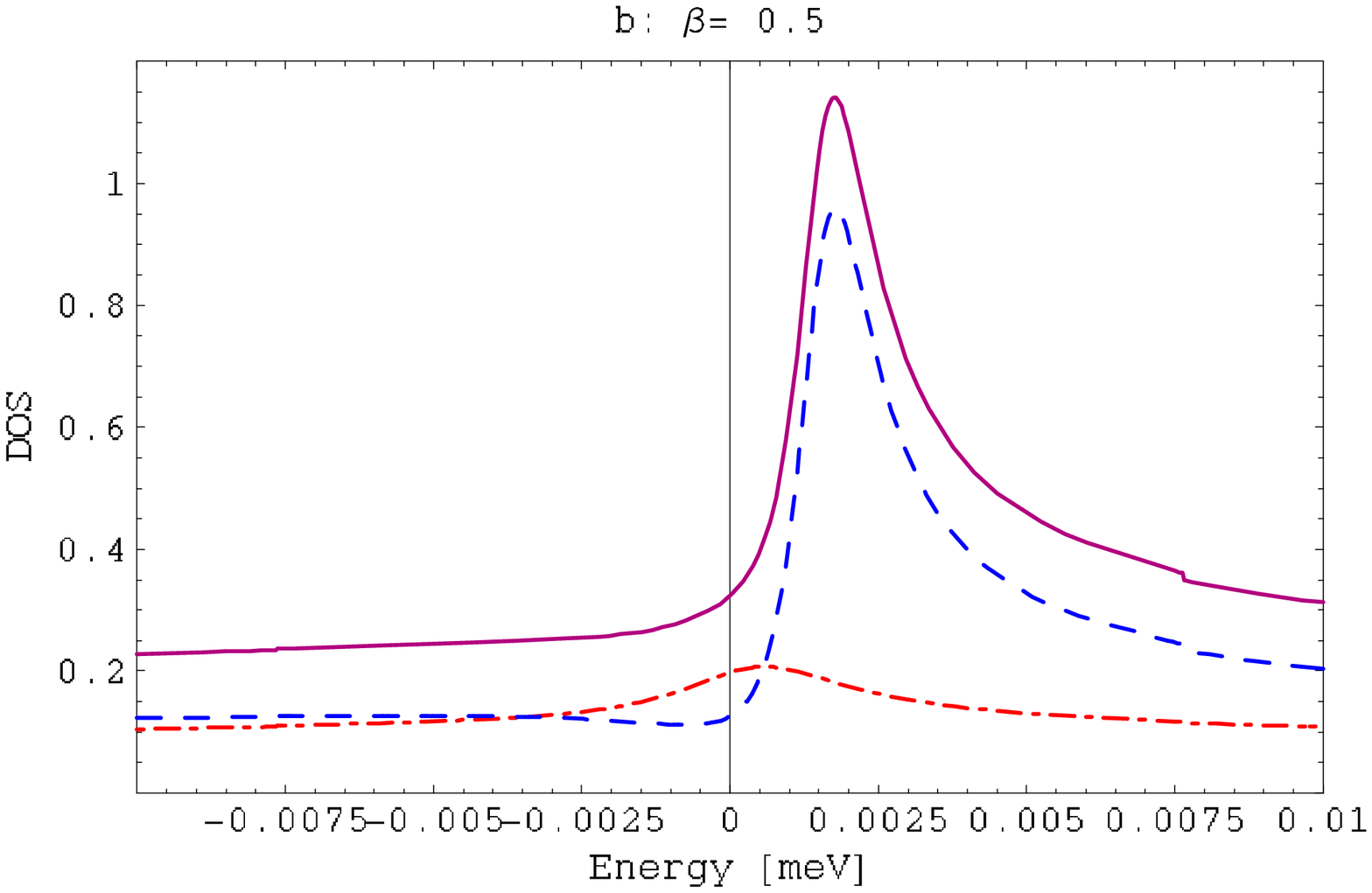}} \hspace{-0.6 cm} \subfigure[~$\beta=1$]
{\includegraphics[width=.33\textwidth]{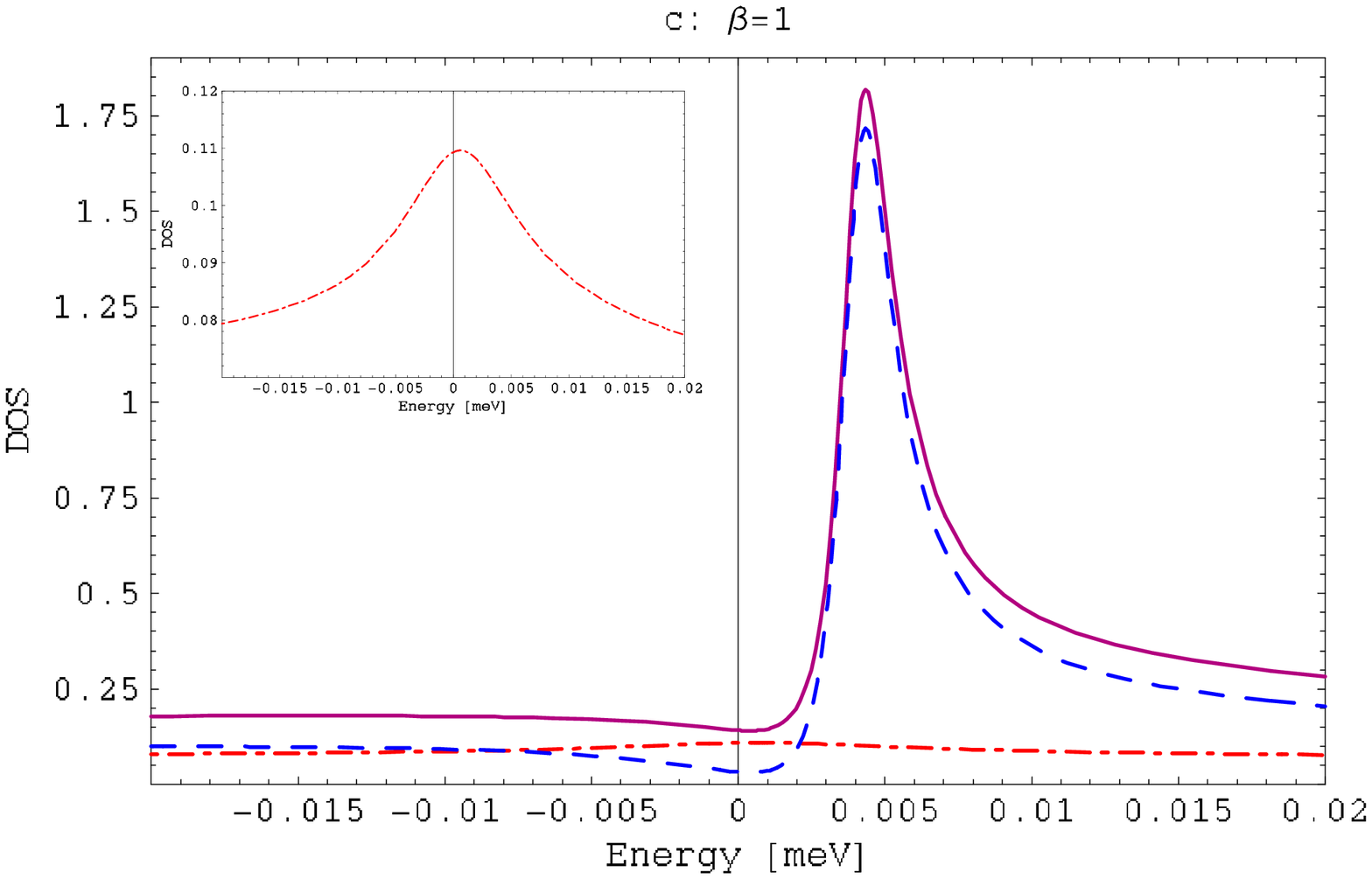}}
\caption{{\protect\small {We plot the DOS with various $\protect\beta $
values. Other parameters are $\Delta=0$, $\Gamma =0.2~\mathrm{meV}$, $D=10~%
\mathrm{meV}$, and $\protect\varepsilon _{d}=-2~\mathrm{meV}$. In each
graph, the red dot-dashed curve is plotted from the first term of Eq.~%
\protect\ref{eq:gref1_1}, corresponding to $T_{K2}$, and the blue dashed
curve from the second term, corresponding to $T_{K1}$. The purple solid
curve is a combination of the two. (a) $\protect\beta =0$, $T/T_{K1}(\protect%
\beta =0)=0.6$, and $T_{K1}(\protect\beta =0)=2.83\times 10^{-4}~\mathrm{meV}
$. The red dot-dashed curve overlaps with the blue dashed curve as $%
T_{K1}=T_{K2}$ with the peaks located at $T_{K1}(\protect\beta =0)$.
(b)[(c)] $\protect\beta =0.5[1]$, $T/T_{K1}(\protect\beta =0.5[1])=0.6$. The
blue dashed curve shifts right away from the Fermi level and its peak sits
at $T_{K1}(\protect\beta =0.5[1])=1.26\times 10^{-3}[3.88\times 10^{-3}]~%
\mathrm{meV}$, while the red dot-dashed curve shifts left towards the Fermi
level with its peak located at $T_{K2}(\protect\beta =0.5[1])=3.49\times
10^{-5}[1.51\times 10^{-6}]~\mathrm{meV}$. The inset in (c) shows the peak
corresponding to $T_{K2}(\protect\beta=1)$. Note that the blue dashed curve
is more pronounced than the red dot-dashed curve because $T_{K1}>T_{K2}$ as $%
\protect\beta >0$.}}}
\label{fig:vary_coupling}
\end{figure}

%\begin{figure}[b]
%\epsfig{figure=vary_coupling.eps,width=4.5 in}
%\caption{{\protect\small {At $T/T_{K1}(\protect\beta=1)=0.6$ and $T_{K1}(%
%\protect\beta=1)=0.004~ \mathrm{meV}$, we compute curves for different
%values $\protect\beta=0(purple),~0.25(red),~0.5(green), ~0.75(dark
%~green),~1(blue)$. As $\protect\beta$ increases, the peak becomes sharper,
%corresponding to increase of $T_{K1}(\protect\beta)$}.}}
%\label{fig:vary_coupling}
%\end{figure}
This enhancement of the Kondo peak and hence the Kondo temperature is caused
by interference between two tunnelings, one with valley index conserved, and
the other without. \ This is evident from the fact that the $T_{K}$'s depend
on the relative sign of $V_{0}$ and $V_{X}.$ \ Constructive interference
increases the Kondo temperature whereas destructive interference decreases
it. This $\beta $-dependent Kondo temperature is somewhat different from
that in \cite{JSLim2006}, where the authors use slave-boson mean field
theory. It would be interesting to investigate the difference between these
two approaches.
%An interesting question:  what is Kondo temperature when valley states $>2$ when valley number is not conserved? Is there any physical system like this?
It should be mentioned that $T_{K}(\beta =0)$ acquired by the EOM approach
underestimates the true Kondo temperature $T_{K}^{\ast }$ which is
\[
k_{B}T_{K}^{\ast }=D\exp {(\pi \varepsilon _{m\sigma }/4\Gamma )}
\]%
This is certainly a deficiency of the EOM approach, and it has been
discussed in the literature \cite{Lacroix1981}. The underestimation of the
Kondo temperature from the EOM approach leads to its failure in determining,
for example, the local spin susceptibility in the Kondo regime \cite%
{Kashcheyevs2006}. However, as we see in the following, the disappearance of
Kondo peak structure suppressed by the temperature and the magnetic field is
only scaled by the Kondo temperature $T_{K1(2)}(\beta )$ and hence the EOM
approach faithfully captures the Kondo effect.

\section{V. Density of States}

In Fig. 1 we showed the level structure of a single orbital on the QD. \ The
four energy states suffer both Zeeman splitting and valley splitting. \ We
now formally define these energies as $\varepsilon _{m\sigma }=\varepsilon
_{d}+(\Delta /2+\mu _{v}B)(\delta _{m,e}-\delta _{m,o})+g\mu _{B}B(\delta
_{\sigma ,\uparrow }-\delta _{\sigma ,\downarrow })$, where $\varepsilon
_{d} $ is the bare energy of the dot and $B$ is the applied magnetic field. $%
\Delta $ is the zero-field valley splitting. Experiments in $2$-DEGs also
show that the valley splitting increases linearly with the magnetic field
\cite{srijit_nature}. For a QD the valley splitting slope depends on the
size of the dot. A small Si QD can have $\mu _{v}$ relatively comparable to $%
g\mu _{B}\sim 0.1~\mathrm{meV/T}$, with $g=2$.

In the following, we consider four different situations with different
values of the three parameters $B,~\beta ,~\Delta $ and in each discuss the
peak structure in the DOS in full detail. \ We have chosen the parameters $%
\Gamma =0.2$ meV, $D=10$ meV and $\varepsilon _{d}=-2$ meV. \ This leads to
rather low Kondo temperatures, but is favorable for illustration purposes.

\subsection{A. $B=0$ and $\Delta =0$.}

This is the case considered in the previous section. \ $\Delta =0$ is
probably not achievable in experiments, but this case is still important to
understand, since it allows us to isolate the effect of valley index
non-conservation. We showed in the previous section that there are two $%
\beta $-dependent Kondo peaks coming from the two terms of Eq.~\ref%
{eq:gref1_1}, the first term corresponding to $T_{K2}$, the second term
corresponding to $T_{K1}$. \ \ The DOS\ can be similarly partitioned, and we
plot each piece separately together with their sum. \ Fig.~\ref%
{fig:vary_coupling} shows the Kondo peak structure at $\beta =0,~0.5,~1$. We
plot each graph at temperature $T=0.6~T_{K1}(\beta )$. \ When $\beta =0$
(valley index conserved), $T_{K1}=T_{K2}$, the two Kondo peaks overlap with
the peak position at $T_{K1}$, each contributing to half of the total
weight. As $\beta $ increases and tunnelings between different valley states
become allowed, $T_{K1}(\beta )$ increases with $\beta $, while $%
T_{K2}(\beta )$ decreases, according to Eq.~\ref{eq:kondoTK1} and \ref%
{eq:kondoTK2}. Since we set the temperature $T$ slightly smaller than $%
T_{K1}(\beta )$, but much larger than $T_{K2}(\beta =0.5,1)$, we expect to
see one pronounced peak shift left and sit at energy $T_{K1}(\beta )$, and
the other much suppressed peak shifts right towards $T_{K2}(\beta )$, as
shown in Fig.~\ref{fig:vary_coupling}. Therefore, when tunnelings between
different valley states are allowed, it will enhance exponentially one Kondo
peak with the Kondo temperature $T_{K1}$, and suppress exponentially the
other Kondo peak with the Kondo temperature $T_{K2}$. \ \ $T_{K1}$
overshadows $T_{K2}$ except in the neighborhood of $\beta =0.5$, where two
peaks are relatively well separated and both sizable. \ As stated above, it
is probably difficult to achieve $\Delta =0$ experimentally. \ However, even
if $\Delta $ is small, we expect an enhancement of the upper Kondo
temperature, making the features in the DOS associated with this $T_{K}$
easier to observe. \ In this sense, Si is more favorable than GaAs for
observation of the Kondo effect.

\subsection{B. $B=0$, $\protect\beta =0$.} 
\begin{figure}[t]
\begin{center}
\setlength{\unitlength}{1cm}
\begin{picture}(5,5)(2,1)
 \includegraphics[width=0.7\linewidth,clip]{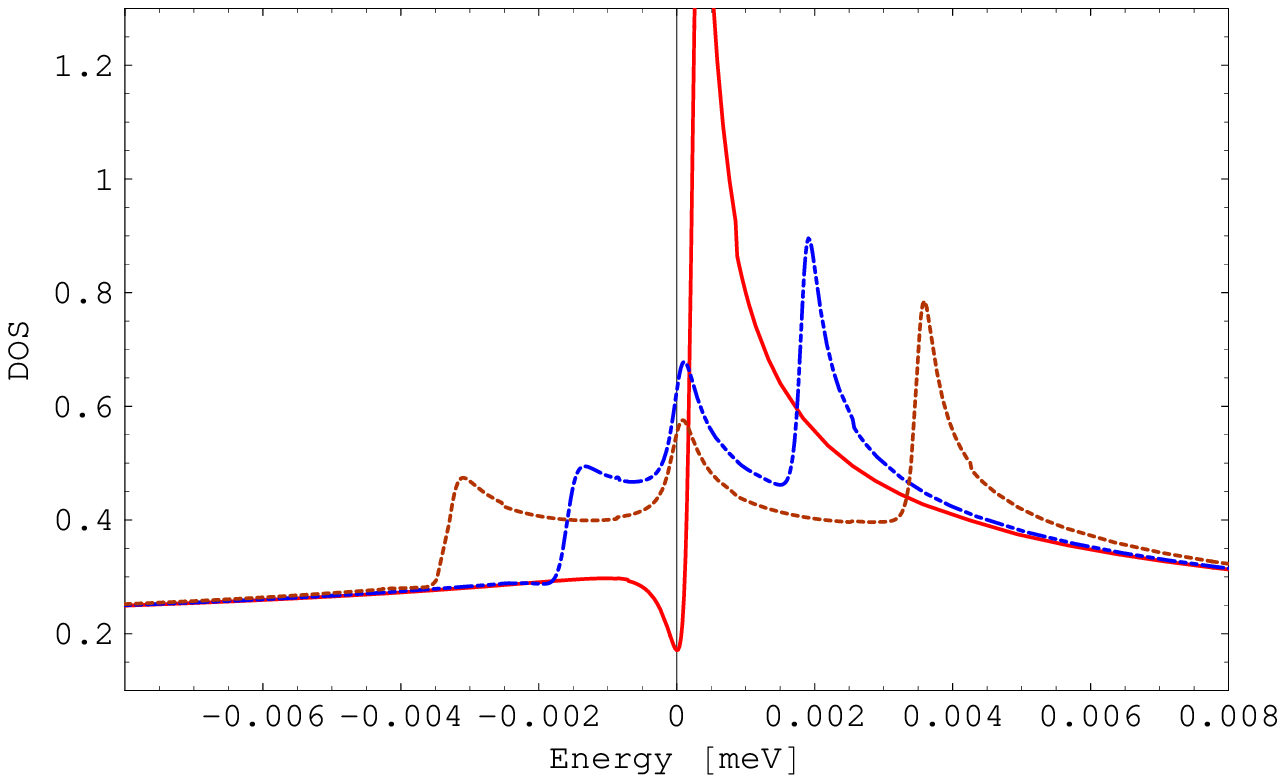}
     \end{picture}
\begin{picture}(5,5)(5,-2.5)
\includegraphics[width=0.5\linewidth,clip]{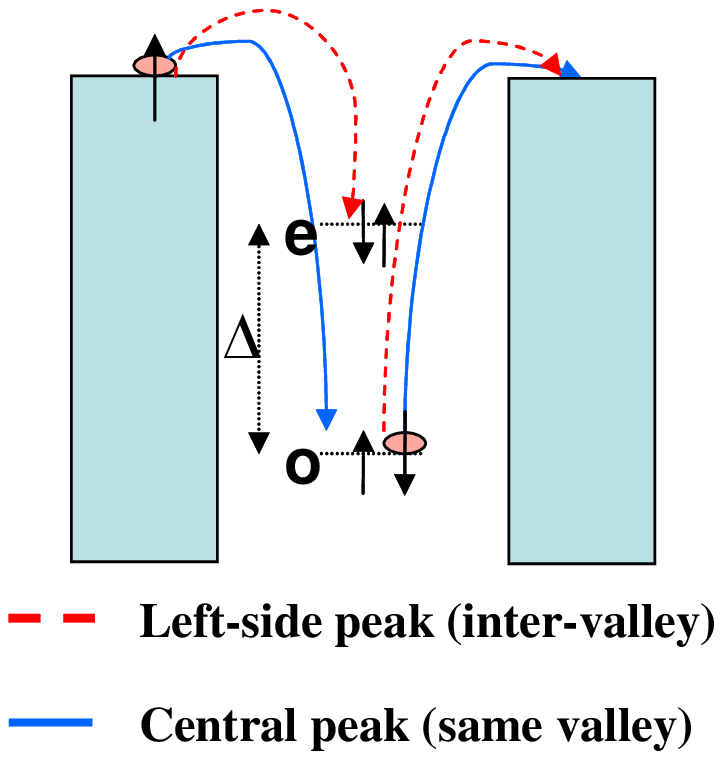}
  \end{picture}
\end{center}
\caption{{\protect\small {\ We plot the DOS with $\protect\beta =0$, $B=0$
and $\Delta =0$(red solid curve), }}$\Delta =${\protect\small {$6~T_{K1}$%
(blue dot-dashed curve), and }}$\Delta =${\protect\small {$12~T_{K1}$(brown
dotted curve) at a temperature $T=0.3~T_{K1}$ where $T_{K1}=2.83\times
10^{-4}~\mathrm{meV}$. The other parameters are the same as in Fig.~ \protect
\ref{fig:vary_coupling}. Note that at $\Delta =0$, the peak shifts slightly
away from the Fermi level and locates itself at $T_{K1}$. As $\Delta $
increases, the peak splits into three peaks at the positions $-\Delta
,~0,~\Delta $. The central peak comes from spin exchange at the same valley
state (no energy penalty) and shifts back to the Fermi level, as expected
from the conventional spin-$1/2$ Kondo effect. The other two side peaks
result from inter-valley processes (energy cost $\Delta $). A schematic on
the top left depicts two transitions that contribute to the central peak
and left-side peak.}}}
\label{fig:kondo_delta}
\end{figure}

\begin{figure}[t]
\epsfig{figure=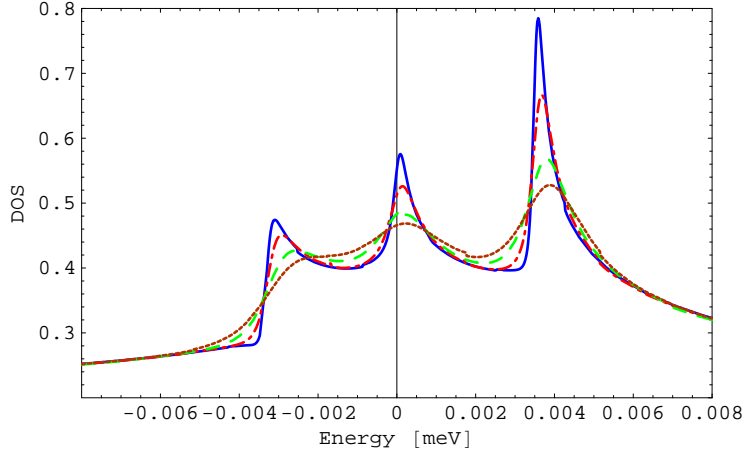,width=0.6\linewidth}
\caption{{\protect\small {We plot the DOS with $\protect\beta =0,~\Delta
=12~T_{K1}$ at various temperatures: $T/T_{K1}=0.3$ (blue solid curve) , $%
T/T_{K1}=$$0.6$ (red dot-dashed curve), $T/T_{K1}=$$1.2$ (green dashed
curve), and $T/T_{K1}=$ $1.8$ (brown dotted curve). The Kondo temperature $%
T_{K1}=2.84\times 10^{-4}~\mathrm{meV}$. The other parameters are the same
as in Fig.~\protect\ref{fig:vary_coupling}. Here the three Kondo peaks are
equally suppressed by increase of the temperature.}}}
\label{fig:kondo_T}
\end{figure}

We now consider what happens when there is a finite valley splitting at zero
field. \ We find that there are three peaks in the DOS\ (and hence in the
nonlinear conductance) instead of the usual two. \ There is a central peak
at zero energy and side peaks at $\pm \Delta .$\ \ This is shown in Fig.~\ref%
{fig:kondo_delta}. \ \ The schematic diagram on the top left shows the
conduction processes corresponding to the central peak and and the left-side
peak. As clearly demonstrated in the schematic, there are basically two
energetically distinguishable types of transitions, the inter-valley
transition and the spin-flip transition at the same valley. Because valley
degeneracy is already broken with a splitting $\Delta $, the inter-valley
transitions cost an energy penalty $\Delta $, therefore shifting the peaks
away from the Fermi level to $\pm \Delta $. They represent the transitions
where an electron in the conduction bands enters into one valley state while
an electron hops out of the QD from the other valley state (red dotted curve
in the schematic). Furthermore, there is no energy cost in the spin-flip
transition at the same valley, since spin degeneracy is still preserved, so
the corresponding peak locates itself at the Fermi level (blue solid curve
in the schematic). This is basically the same scenario as traditional spin-$%
1/2$ GaAs QDs. Fig.~\ref{fig:kondo_T} shows that increase of temperature
suppresses these three Kondo peaks equally, demonstrating that they are
scaled by the same Kondo temperature.

We further calculate the Kondo temperature with $\Delta \not=0$. Since $%
\beta =0$, we reduce the real part of the denominator of $\mathcal{G}%
_{m\sigma }$ to
\begin{equation}
w-\varepsilon _{m\sigma }-\sum_{l\not=m\sigma }\Sigma _{c,l}(z)=0
\label{cal_tk}
\end{equation}%
Consider $"m\sigma "="o\uparrow "$ energy state with the energy $\varepsilon
_{o,\uparrow }=\varepsilon _{d}-\Delta /2$. As $T\rightarrow 0$, Eq.~\ref%
{cal_tk} can be rewritten as
\[
w-\varepsilon _{d}+\Delta /2+\Gamma \mathrm{ln}\left\vert \frac{w}{D}%
\right\vert +2\Gamma \mathrm{ln}\left\vert \frac{w+\Delta }{D}\right\vert =0%
\nonumber
\]%
Again we set $w=k_{B}T_{K}(\Delta )$. In the region $k_{B}T_{K}(\Delta
=0)\ll |\Delta |\ll D$, because $T_{K}(\Delta \not=0)<T_{K}(\Delta =0)$, $%
k_{B}T_{K}(\Delta \not=0)\ll \Delta $. Therefore, we neglect linear $w$ and
the last logarithmic term $2\Gamma \mathrm{ln}\left\vert \frac{w+\Delta }{D}%
\right\vert \sim 2\Gamma \mathrm{ln}\left\vert \frac{\Delta }{D}\right\vert $%
. We obtain
\begin{eqnarray}
k_{B}T_{K}(\Delta ) &=&D\exp ((\varepsilon _{d}-\Delta /2)/\Gamma -2\mathrm{%
ln}\left\vert \Delta /D\right\vert )  \nonumber \\
&=&\frac{(k_{B}T_{K}(\Delta =0))^{3}}{\Delta ^{2}}  \label{eom_tk}
\end{eqnarray}%
Thus using EOM, we find $T_{K}(\Delta )\sim 1/\Delta ^{2}$. Eto in his paper
\cite{Eto2005} evaluates theoretically the Kondo temperature in the QDs with
two orbitals and spin-$1/2$ as a function of $\Delta $, the energy
difference between the two orbitals. He considers the case $\beta =0$ so the
two Kondo temperatures coincide. \ Using "poor man's" scaling method, he
finds that within the region where $T_{K}(\Delta =0)\ll |\Delta |\ll D$, $%
T_{K}(\Delta )$ decreases as $\Delta $ increases, according to a power law
\[
k_{B}T_{K}(\Delta )=k_{B}T_{K}(\Delta =0)\cdot (k_{B}T_{K}(\Delta
=0)/|\Delta |)^{\gamma } \nonumber
\]%
where $\gamma =1$. The discrepancy of Eq.~\ref{eom_tk} from Eto's power law
is again due to the deficiency of EOM approach to find the true Kondo
temperature.

Interestingly, as $\Delta $ becomes larger, the side peaks shift further
away from the Fermi level and we recover the conventional spin-$1/2$ Kondo
effect, with one Kondo peak at the Fermi level. This is in agreement with
Eto's analysis \cite{Eto2005}.

\begin{figure}[tbp]
\vspace{-4 cm}
\par
\begin{center}
\begin{minipage}[t]{0.48\linewidth}
 \epsfig{file=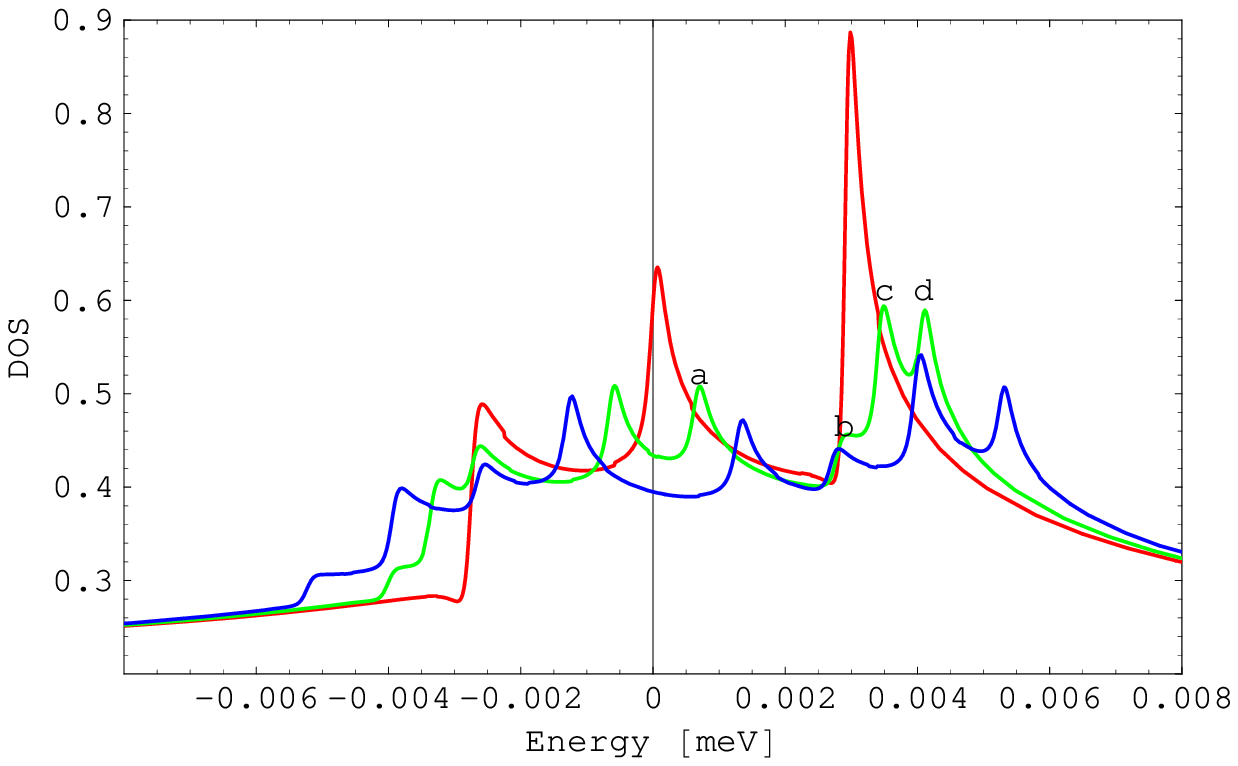,clip=, width=\linewidth}
    \end{minipage}\hfill
\begin{minipage}[t]{0.48\linewidth}
\epsfig{file=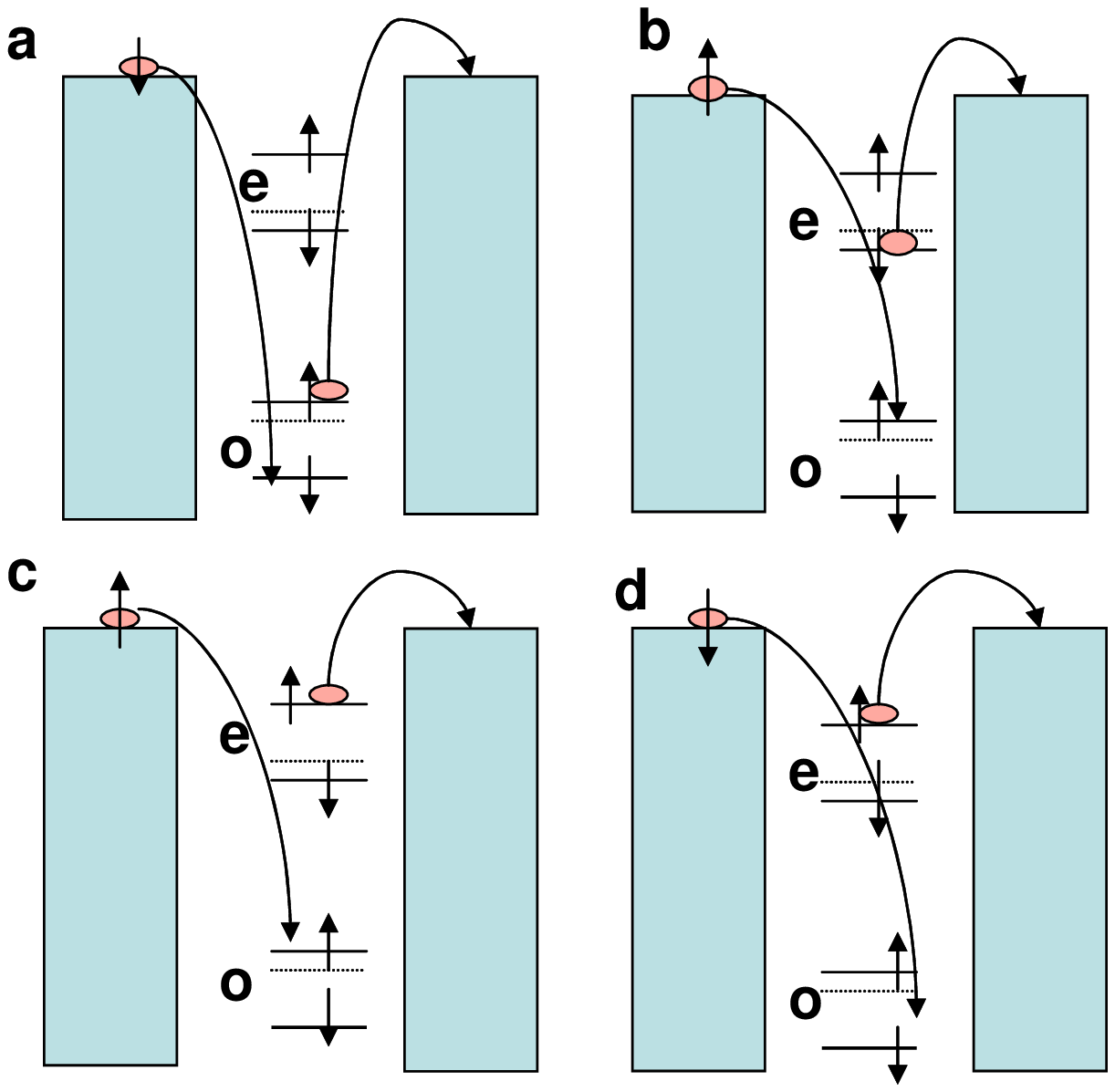,clip=, width=\linewidth}
    \end{minipage}
\end{center}
\caption{{\protect\small {The DOS is plotted at $\protect\beta =0$ with the
magnetic field $B=0$ (red curve), $0.00284~\mathrm{T}$ (green curve) and $%
0.00568~\mathrm{T}$ (blue curve). The Kondo temperature $T_{K1}=2.84\times
10^{-4}~\mathrm{meV}$. We set $T/T_{K1}=0.2$. The other parameters are the
same as in Fig.~\protect\ref{fig:vary_coupling}. We set $\protect\mu %
_{v}=0.1~\mathrm{meV/T}$, slightly smaller than $g\protect\mu _{B}=0.114~%
\mathrm{meV/T}$. The central peak splits into two peaks because the spin
degeneracy is broken, while three energetically distinguishable inter-valley
transitions due to the presence of magnetic field split the side peak into
three peaks. The transitions that result in peaks $a,~b,~c,~d$(green curve)
are shown in the schematic diagrams $a,~b,~c,~d$, respectively. Moreover,
each peak evolves with a different slope with respect to the magnetic field
due to cancellation or addition of spin and valley slopes. The slopes of
peaks $a,~b,~c,~d$ are $0.228,~-0.028,~0.2,~0.428~\mathrm{meV/T}$,
respectively.}}}
\label{fig:kondo_B}
\end{figure}

\subsection{C. $\protect\beta =0$.}

This is now the more general case of finite field and valley splitting, but
still assuming that valley index is conserved. \ Now the peak structure in
the DOS becomes more complicated. As already seen in Fig.~\ref%
{fig:kondo_delta}, at $B=0~$and $\Delta \not=0$, there are three peaks at $%
w=0,~\pm \Delta $. By applying a magnetic field, hence breaking spin
degeneracy, as shown in Fig.~\ref{fig:kondo_B}, the central peak splits into
two peaks with peak positions at $\pm 2g\mu _{B}B$ by Zeeman effect. In Fig.~%
\ref{fig:kondo_B}, the schematic diagram $a$ demonstrates a spin-flip
transition that contributes to the peak at position $2g\mu _{B}B$. The side
peak above the Fermi level splits into three peaks with peak positions at $%
\Delta +2(\mu _{v}-g\mu _{B})B,~\Delta +2\mu _{v}B,~\Delta +2(\mu
_{v}+g\mu _{B})B$ which correspond to inter-valley transitions in
the schematic diagrams $b,~c,~d$, respectively, whereas the side
peak below the Fermi level also splits into three peaks with peak
positions at $-\Delta -2(\mu _{v}+g\mu _{B})B,~-\Delta -2\mu
_{v}B,~-\Delta -2(\mu _{v}-g\mu _{B})B$ by the same token. It is
easy to see that twice the Zeeman energy $2g\mu _{B}B$ corresponds
to the energy cost for spin-flip transitions, while valley
splitting energy $\Delta +2\mu _{v}B$ corresponds to that for
inter-valley transitions. The positions of these eight Kondo peaks
have different dependencies on the magnetic field, with their
slopes being expressed as various linear combinations of $\Delta
,~\mu _{v}$, and $~g\mu _{B}$. It is worth noting that equally
weighted peaks at positions $\pm \Delta \pm 2(\mu _{v}+g\mu
_{B})B,~\pm \Delta \pm 2(\mu _{v}-g\mu _{B})B$ are twice as small
as the other peaks because only half the number of conduction
processes contribute to the peaks.

\begin{figure}[t]
\begin{center}
\setlength{\unitlength}{1cm}
\begin{picture}(5,5)(1,1)
 \includegraphics[width=0.7\linewidth,clip]{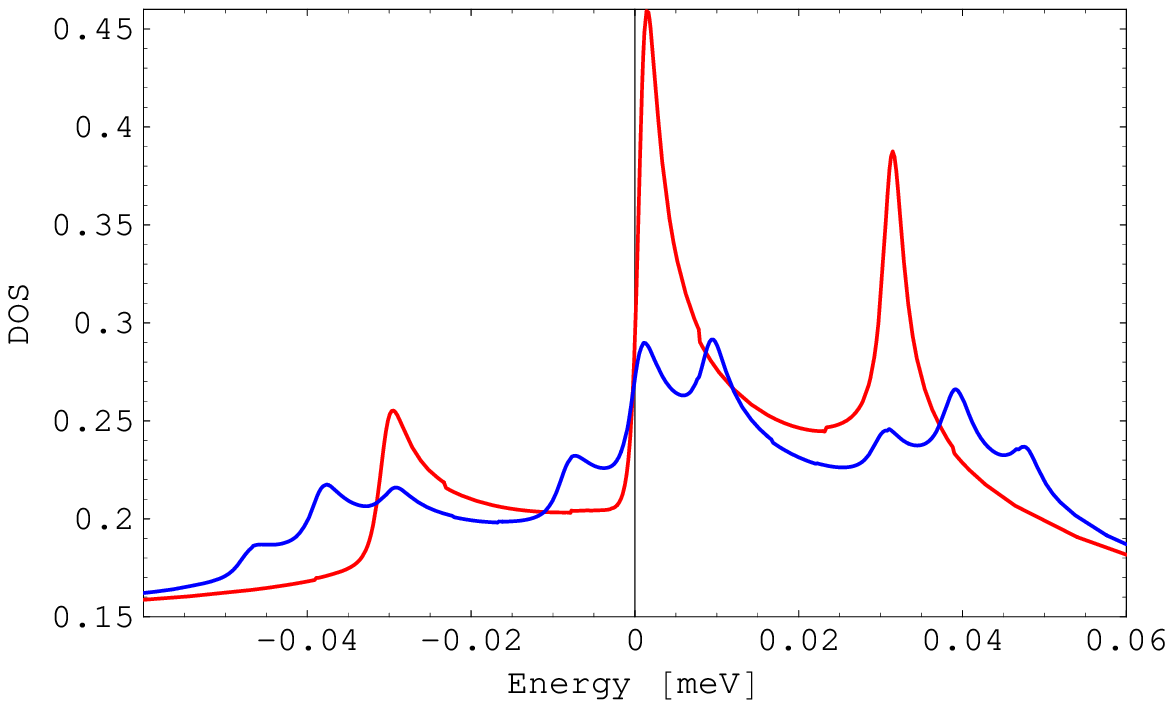}
     \end{picture}
\begin{picture}(5,5)(-3,-2.5)
\includegraphics[width=0.6\linewidth,clip]{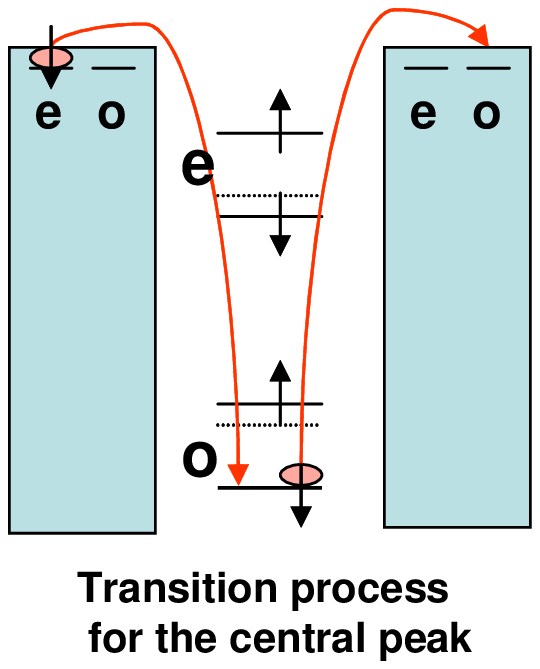}
  \end{picture}
\end{center}
\caption{{\protect\small {The Kondo peaks are shown with $B=0~T$ (red curve)
and $B=0.0388~T$ (blue curve) when the valley index is not conserved. We
consider $\protect\beta =1$, $\Delta =8~T_{K1}~\mathrm{meV}$ and the Kondo
temperature $T_{K1}=3.88\times 10^{-3}~\mathrm{meV}$. We set $T/T_{K1}=0.2$.
The other parameters are the same as in Fig.~\protect\ref{fig:vary_coupling}%
. Note that there remains a peak at the Fermi level even when the magnetic
field is not zero, in contrast to the peak structure when valley index is
conserved (cf. Fig.~\protect\ref{fig:kondo_B}). This peak comes from a
transition where a conduction electron from an even valley state hops onto
the dot at the odd valley state while the electron at the odd valley state
hops out of the dot to a lead at an odd valley state, as shown in the top
right schematic. This kind of transition costs no energy penalty.}}}
\label{fig:valley_conser_B}
\end{figure}

\subsection{D. $\protect\beta =1$.}

This is the case where tunneling between same valley states and different
valley states are equally strong. \ When the magnetic field is zero, there
are three peaks near the Fermi level which we have explained in subsection {\bf B}. 
However, as can be seen in Fig.~\ref{fig:valley_conser_B}, the central
peak is more pronounced than the other two side peaks. This is due to extra
contribution from tunnelings between different valley states. The effect of
non-conservation of valley index can be seen more clearly when we apply a
magnetic field. When $B\not=0$ and assuming the zero-field valley splitting $%
\Delta \not=0$, we saw eight peaks since spin and valley states are no
longer degenerate. Surprisingly, as shown in Fig.~\ref{fig:valley_conser_B},
when the valley index is no longer conserved, we not only see the previous
eight peaks as in Fig.~\ref{fig:kondo_B} when the valley index is conserved,
but also see an extra peak at the Fermi level that is independent of the
magnetic field, meaning it can only be suppressed by increasing the
temperature. This peak comes from a logarithmic divergence in the function $%
\Sigma _{d}(w)$ (cf. Eq.~\ref{fn:sma_d}) and is thus more pronounced as $%
\beta $ becomes larger. It also indicates that through cross-valley
couplings $V_{X}$, $"m\sigma "$ and $"\bar{m}\sigma "$ form a bound state
through exchanging electrons with the leads. The schematic in Fig.~\ref%
{fig:valley_conser_B} demonstrates that this kind of transition through
cross-valley couplings can occur without any energy cost. Consequently its
peak sits at the Fermi level and has no dependence on the magnetic field. \
This is in some sense the most characteristic signature that could be sought
of the valley Kondo effect, since it results from a bound state of different
valley states rather than different spin states.

\section{VI. Conclusions and implications for experiments}

Silicon quantum dots show a much richer phenomenology than GaAs dots due to
the additional valley degree of freedom. \ The initially fourfold degenerate
level is split at zero applied magnetic field by interface effects and there
is additional splitting at finite fields. \ We have computed the density of
states on the dot in the large-$U$ limit using an equation of motion method
that has been shown to give reasonable results in related problems. \ This
function should closely mimic the non-linear conductance through the dot as
a function of voltage.

\ The nonlinear conductance near zero bias of a Si QD with a single electron
is sharply different from that of a GaAs QD in the Kondo regime:

1. At $B=0$ we expect three peaks near zero bias at $e\mathrm{V}=0,\pm
\Delta ,$ where $\Delta $ is the valley splitting. \ The central peak is the
spin Kondo peak while the other peaks correspond to inter-valley conduction
processes.

2. For $B>0,$ the side peaks further split into three peaks, the two
split-off peaks corresponding to spin-flip processes and the middle
corresponding to spin conserving processes.

3. For $B>0,$ the central peak splits into two peaks if the valley index is
conserved and three if it is not conserved, the latter corresponding to a
valley-Kondo ground state.

4. The temperature dependence can be complex, since there are two Kondo
temperatures in the problem. \

5. The conservation or non-conservation of valley index depends quite
sensitively on the precise details of the junctions and is therefore likely
to be sample-dependent and difficult to control.

We hope that some or all of these phenomena will be observed in future
experiments.

\section{Acknowledgements}

We gratefully acknowledge conversations with M. Friesen. \ This work was
supported by NSA and ARDA under ARO contract number W911NF-04-1-0389 and by
the National Science Foundation through the ITR (DMR-0325634) and EMT
(CCF-0523675) programs.

\renewcommand{\theequation}{\mbox{A.\arabic{equation}}} %\section{Appendix}
\setcounter{equation}{0} % reset counter

\section*{APPENDIX: EOM FOR INFINITE $U$}

The basic equations of motion for the Green's functions are
\begin{eqnarray}
(w-\varepsilon _{m\sigma })\ll f_{m\sigma },f_{m\sigma }^{+}\gg
&=&1+\sum_{ik}V_{0,ik}\ll c_{ikm\sigma },f_{m\sigma }^{+}\gg
+\sum_{ik}V_{X,ik}\ll c_{ik\bar{m}\sigma },f_{m\sigma }^{+}\gg  \nonumber \\
&&+U\sum_{l\not=m\sigma }\ll f_{l}^{+}f_{l}f_{m\sigma },f_{m\sigma }^{+}\gg .
\nonumber \\
(w-\varepsilon _{k})\ll c_{ikm\sigma },f_{m\sigma }^{+}\gg &=&V_{0,ik}\ll
f_{m\sigma },f_{m\sigma }^{+}\gg +V_{X,ik}\ll f_{\bar{m}\sigma },f_{m\sigma
}^{+}\gg .  \nonumber \\
(w-\varepsilon _{\bar{m}\sigma })\ll f_{\bar{m}\sigma },f_{m\sigma }^{+}\gg
&=&\sum_{ik}V_{0,ik}\ll c_{ik\bar{m}\sigma },f_{m\sigma }^{+}\gg
+\sum_{ik}V_{X,ik}\ll c_{ikm\sigma },f_{m\sigma }^{+}\gg  \nonumber \\
&&+U\sum_{p\not=\bar{m}\sigma }\ll f_{p}^{+}f_{p}f_{\bar{m}\sigma
},f_{m\sigma }^{+}\gg .  \nonumber \\
(w-\varepsilon _{k})\ll c_{ik\bar{m}\sigma },f_{m\sigma }^{+}\gg
&=&V_{0,ik}\ll f_{\bar{m}\sigma },f_{m\sigma }^{+}\gg +V_{X,ik}\ll
f_{m\sigma },f_{m\sigma }^{+}\gg .  \nonumber
\end{eqnarray}%
Note that $l,p$ are shorthand for valley plus spin indices, unlike $m$ that
means only the valley index. Furthermore if $l=m\sigma $, $\bar{l}=\bar{m}%
\sigma $. The \textit{bar} means exchanging only the valley index. From now
on we simplify the calculation by changing $V_{0(X),ik}\rightarrow V_{0(X)}$%
. From the above four coupled equations of motion we can obtain
\begin{eqnarray}
(w-\varepsilon _{m\sigma }-\sum_{ik}\frac{V_{0}^{2}+V_{X}^{2}}{w-\varepsilon
_{k}})\ll f_{m\sigma },f_{m\sigma }^{+}\gg &=&1+\sum_{ik}\frac{2V_{0}V_{X}}{%
w-\varepsilon _{k}}\ll f_{\bar{m}\sigma },f_{m\sigma }^{+}\gg  \nonumber \\
&& +U\sum_{l\not=m\sigma }\ll f_{l}^{+}f_{l}f_{m\sigma },f_{m\sigma }^{+}\gg
\label{eq:aa1} \\
(w-\varepsilon _{\bar{m}\sigma }-\sum_{ik}\frac{V_{0}^{2}+V_{X}^{2}}{%
w-\varepsilon _{k}})\ll f_{\bar{m}\sigma },f_{m\sigma }^{+}\gg &=&\sum_{ik}%
\frac{2V_{0}V_{X}}{w-\varepsilon _{k}}\ll f_{m\sigma },f_{m\sigma }^{+}\gg
\nonumber \\
&& +U\sum_{p\not=\bar{m}\sigma }\ll f_{p}^{+}f_{p}f_{\bar{m}\sigma
},f_{m\sigma }^{+}\gg  \label{eq:aa2}
\end{eqnarray}%
The higher-order terms give rise to
\begin{eqnarray}
(w-\varepsilon _{m\sigma }-U)\ll n_{l}f_{m\sigma },f_{m\sigma }^{+}\gg
&=&<n_{l}>+\sum_{ik}V_{0}\{\ll n_{l}c_{ikm\sigma },f_{m\sigma }^{+}\gg +\ll
f_{l}^{+}c_{ikl}f_{m\sigma },f_{m\sigma }^{+}\gg  \nonumber \\
&&-\ll c_{ikl}^{+}f_{l}f_{m\sigma },f_{m\sigma }^{+}\gg
\}+\sum_{ik}V_{X}\{\ll n_{l}c_{ik\bar{m}\sigma },f_{m\sigma }^{+}\gg
\nonumber \\
&&+\ll f_{l}^{+}c_{ik\bar{l}}f_{m\sigma },f_{m\sigma }^{+}\gg -\ll c_{ik\bar{%
l}}^{+}f_{l}f_{m\sigma },f_{m\sigma }^{+}\gg \}  \nonumber \\
&&+U\sum_{j\not=l\not=m\sigma }\ll n_{l}n_{j}f_{m\sigma },f_{m\sigma }^{+}\gg  \label{eq:f4}
\\
(w-\varepsilon _{\bar{m}\sigma }-U)\ll n_{p}f_{\bar{m}\sigma },f_{m\sigma
}^{+}\gg &=&-\delta _{p,m\sigma }<f_{m\sigma }^{+}f_{\bar{m}\sigma
}>+\sum_{ik}V_{0}\{\ll n_{p}c_{ik\bar{m}\sigma },f_{m\sigma }^{+}\gg
\nonumber \\
&&+\ll f_{p}^{+}c_{ikp}f_{\bar{m}\sigma },f_{m\sigma }^{+}\gg -\ll
c_{ikp}^{+}f_{p}f_{\bar{m}\sigma },f_{m\sigma }^{+}\gg \}  \nonumber \\
&&+\sum_{ik}V_{X}\{\ll n_{p}c_{ikm\sigma },f_{m\sigma }^{+}\gg +\ll
f_{p}^{+}c_{ik\bar{p}}f_{\bar{m}\sigma },f_{m\sigma }^{+}\gg  \nonumber \\
&&-\ll c_{ik\bar{p}}^{+}f_{p}f_{\bar{m}\sigma },f_{m\sigma }^{+}\gg
\}+U \sum_{j\not=p\not=\bar{m}\sigma }\ll n_{j}n_{p}f_{\bar{m}\sigma
},f_{m\sigma }^{+}\gg  \label{eq:ff4}
\end{eqnarray}%
%
%
%
%
%The last term is approximately equal to $\frac{U}{w-\varepsilon_m-2U}\sum_{j\not=l\not=m}<n_ln_j> \sim -\frac{N-2}{2}\frac{1}{U} \sim 0$ as $U \rightarrow \infty$.

Next, we proceed to obtain equations of motion for the higher-order
functions appearing on the right-hand side of Eq.~\ref{eq:f4} and Eq.~\ref%
{eq:ff4}
\begin{eqnarray}
(w-\varepsilon _{k})\ll n_{l}c_{ikm\sigma },f_{m\sigma }^{+}\gg  &=&-\delta
_{l,m\sigma }<f_{m\sigma }^{+}c_{ikm\sigma }>+V_{0}\{\ll n_{l}f_{m\sigma
},f_{m\sigma }^{+}\gg   \nonumber \\
&&-\sum_{i^{\prime }k^{\prime }}\ll c_{i^{\prime }k^{\prime
}l}^{+}f_{l}c_{ikm\sigma },f_{m\sigma }^{+}\gg +\sum_{i^{\prime }k^{\prime
}}\ll f_{l}^{+}c_{i^{\prime }k^{\prime }l}c_{ikm\sigma },f_{m\sigma }^{+}\gg
\}  \nonumber \\
&&+V_{X}\{\ll n_{l}f_{\bar{m}\sigma },f_{m\sigma }^{+}\gg -\sum_{i^{\prime
}k^{\prime }}\ll c_{i^{\prime }k^{\prime }\bar{l}}^{+}f_{l}c_{ikm\sigma
},f_{m\sigma }^{+}\gg   \nonumber \\
&&+\sum_{i^{\prime }k^{\prime }}\ll f_{l}^{+}c_{i^{\prime }k^{\prime }\bar{l}%
}c_{ikm\sigma },f_{m\sigma }^{+}\gg \}  \label{eq:f5} \\
(w-\varepsilon _{k})\ll n_{l}c_{ik\bar{m}\sigma },f_{m\sigma }^{+}\gg
&=&-\delta _{l,m\sigma }<f_{m\sigma }^{+}c_{ik\bar{m}\sigma }>+V_{0}\{\ll
n_{l}f_{\bar{m}\sigma },f_{m\sigma }^{+}\gg   \nonumber \\
&&-\sum_{i^{\prime }k^{\prime }}\ll c_{i^{\prime }k^{\prime }l}^{+}f_{l}c_{ik%
\bar{m}\sigma },f_{m\sigma }^{+}\gg +\sum_{i^{\prime }k^{\prime }}\ll
f_{l}^{+}c_{i^{\prime }k^{\prime }l}c_{ik\bar{m}\sigma },f_{m\sigma }^{+}\gg
\}  \nonumber \\
&&+V_{X}\{\ll n_{l}f_{m\sigma },f_{m\sigma }^{+}\gg -\sum_{i^{\prime
}k^{\prime }}\ll c_{i^{\prime }k^{\prime }\bar{l}}^{+}f_{l}c_{ik\bar{m}%
\sigma },f_{m\sigma }^{+}\gg   \nonumber \\
&&+\sum_{i^{\prime }k^{\prime }}\ll f_{l}^{+}c_{i^{\prime }k^{\prime }\bar{l}%
}c_{ik\bar{m}\sigma },f_{m\sigma }^{+}\gg \}  \label{eq:f6} \\
(z-\varepsilon _{k})\ll f_{l}^{+}c_{ikl}f_{m\sigma },f_{m\sigma }^{+}\gg
&=&<f_{l}^{+}c_{ikl}>+V_{0}\{\ll f_{l}^{+}f_{l}f_{m\sigma },f_{m\sigma
}^{+}\gg -\sum_{i^{\prime }k^{\prime }}\ll c_{i^{\prime }k^{\prime
}l}^{+}c_{ikl}f_{m\sigma },f_{m\sigma }^{+}\gg   \nonumber \\
&&-\sum_{i^{\prime }k^{\prime }}\ll c_{ikl}f_{l}^{+}c_{i^{\prime }k^{\prime
}m\sigma },f_{m\sigma }^{+}\gg \}+V_{X}\{\ll f_{l}^{+}f_{\bar{l}}f_{m\sigma
},f_{m\sigma }^{+}\gg   \nonumber \\
&&-\sum_{i^{\prime }k^{\prime }}\ll c_{i^{\prime }k^{\prime }\bar{l}%
}^{+}c_{ikl}f_{m\sigma },f_{m\sigma }^{+}\gg -\sum_{i^{\prime }k^{\prime
}}\ll c_{ikl}f_{l}^{+}c_{i^{\prime }k^{\prime }\bar{m}\sigma },f_{m\sigma
}^{+}\gg \}  \label{eq:f7} \\
(z-\varepsilon _{k})\ll f_{l}^{+}c_{ik\bar{l}}f_{m\sigma },f_{m\sigma
}^{+}\gg  &=&<f_{l}^{+}c_{ik\bar{l}}>+V_{0}\{\ll f_{l}^{+}f_{\bar{l}%
}f_{m\sigma },f_{m\sigma }^{+}\gg -\sum_{i^{\prime }k^{\prime }}\ll
c_{i^{\prime }k^{\prime }l}^{+}c_{ik\bar{l}}f_{m\sigma },f_{m\sigma }^{+}\gg
\nonumber \\
&&-\sum_{i^{\prime }k^{\prime }}\ll c_{ik\bar{l}}f_{l}^{+}c_{i^{\prime
}k^{\prime }m\sigma },f_{m\sigma }^{+}\gg \}+V_{X}\{\ll
f_{l}^{+}f_{l}f_{m\sigma },f_{m\sigma }^{+}\gg   \nonumber \\
&&-\sum_{i^{\prime }k^{\prime }}\ll c_{i^{\prime }k^{\prime }\bar{l}%
}^{+}c_{ik\bar{l}}f_{m\sigma },f_{m\sigma }^{+}\gg -\sum_{i^{\prime
}k^{\prime }}\ll c_{ik\bar{l}}f_{l}^{+}c_{i^{\prime }k^{\prime }\bar{m}%
\sigma },f_{m\sigma }^{+}\gg \}  \label{eq:f8} \\
\ll c_{ikl}^{+}f_{l}f_{m\sigma },f_{m\sigma }^{+}\gg  &\sim
&1/(w+\varepsilon _{k}-\varepsilon _{l}-\varepsilon _{m\sigma }-U)
\label{eq:f9} \\
\ll c_{ik\bar{l}}^{+}f_{l}f_{m\sigma },f_{m\sigma }^{+}\gg  &\sim
&1/(w+\varepsilon _{k}-\varepsilon _{l}-\varepsilon _{m\sigma }-U)
\label{eq:f10} \\
U\ll n_{l}n_{j}f_{m\sigma },f_{m\sigma }^{+}\gg  &=&U(<n_{l}n_{j}>+\cdots
)/(w-\varepsilon _{m\sigma }-2U)  \label{eq:f11} \\
(\bar{z}-\varepsilon _{k})\ll f_{l}^{+}c_{ikl}f_{\bar{m}\sigma },f_{m\sigma
}^{+}\gg  &=&V_{0}\{\ll f_{l}^{+}f_{l}f_{\bar{m}\sigma },f_{m\sigma }^{+}\gg
-\sum_{i^{\prime }k^{\prime }}\ll c_{i^{\prime }k^{\prime }l}^{+}c_{ikl}f_{%
\bar{m}\sigma },f_{m\sigma }^{+}\gg   \nonumber \\
&&-\sum_{i^{\prime }k^{\prime }}\ll c_{ikl}f_{l}^{+}c_{i^{\prime }k^{\prime }%
\bar{m}\sigma },f_{m\sigma }^{+}\gg \}+V_{X}\{\ll f_{l}^{+}f_{\bar{l}}f_{%
\bar{m}\sigma },f_{m\sigma }^{+}\gg   \nonumber \\
&&-\sum_{i^{\prime }k^{\prime }}\ll c_{i^{\prime }k^{\prime }\bar{l}%
}^{+}c_{ikl}f_{\bar{m}\sigma },f_{m\sigma }^{+}\gg -\sum_{i^{\prime
}k^{\prime }}\ll c_{ikl}f_{l}^{+}c_{i^{\prime }k^{\prime }m\sigma
},f_{m\sigma }^{+}\gg \}  \label{eq:f12} \\
(\bar{z}-\varepsilon _{k})\ll f_{l}^{+}c_{ik\bar{l}}f_{\bar{m}\sigma
},f_{m\sigma }^{+}\gg  &=&V_{0}\{\ll f_{l}^{+}f_{\bar{l}}f_{\bar{m}\sigma
},f_{m\sigma }^{+}\gg -\sum_{i^{\prime }k^{\prime }}\ll c_{i^{\prime
}k^{\prime }l}^{+}c_{ik\bar{l}}f_{\bar{m}\sigma },f_{m\sigma }^{+}\gg
\nonumber \\
&&-\sum_{i^{\prime }k^{\prime }}\ll c_{ik\bar{l}}f_{l}^{+}c_{i^{\prime
}k^{\prime }\bar{m}\sigma },f_{m\sigma }^{+}\gg \}+V_{X}\{\ll
f_{l}^{+}f_{l}f_{\bar{m}\sigma },f_{m\sigma }^{+}\gg   \nonumber \\
&&-\sum_{i^{\prime }k^{\prime }}\ll c_{i^{\prime }k^{\prime }\bar{l}%
}^{+}c_{ik\bar{l}}f_{\bar{m}\sigma },f_{m\sigma }^{+}\gg -\sum_{i^{\prime
}k^{\prime }}\ll c_{ik\bar{l}}f_{l}^{+}c_{i^{\prime }k^{\prime }m\sigma
},f_{m\sigma }^{+}\gg \}  \label{eq:f13} \\
\ll c_{ikl}^{+}f_{l}f_{\bar{m}\sigma },f_{m\sigma }^{+}\gg  &\sim
&1/(w+\varepsilon _{k}-\varepsilon _{l}-\varepsilon _{\bar{m}\sigma }-U)
\label{eq:f14} \\
\ll c_{ik\bar{l}}^{+}f_{l}f_{\bar{m}\sigma },f_{m\sigma }^{+}\gg  &\sim
&1/(w+\varepsilon _{k}-\varepsilon _{l}-\varepsilon _{\bar{m}\sigma }-U)
\label{eq:f15} \\
U\ll n_{l}n_{j}f_{\bar{m}\sigma },f_{m\sigma }^{+}\gg  &=&U(\delta
_{l,m\sigma }<n_{j}f_{m\sigma }^{+}f_{\bar{m}\sigma }>+\cdots
)/(w-\varepsilon _{\bar{m}\sigma }-2U).  \label{eq:f16}
\end{eqnarray}%
%
%
%
%
%, valid for temperatures higher than Kondo temperature, by neglecting terms that involve correlations in the leads (see paper by C. Lacroix) from the above three higher-order equations of motion.
Note $z=w-\varepsilon _{m\sigma }+\varepsilon _{l\not=m\sigma }$ and $\bar{z}%
=w-\varepsilon _{\bar{m}\sigma }+\varepsilon _{l\not=\bar{m}\sigma }$. \ If
we take $U\rightarrow \infty $, Eq.~\ref{eq:f9}-\ref{eq:f11} and Eq.~\ref%
{eq:f14}-\ref{eq:f16} vanish, as do correlators And the rest involving two
annihilation operators on the dot like $\ll f_{l}^{+}f_{\bar{l}}f_{m\sigma
},f_{m\sigma }^{+}\gg $ also vanish as $U\rightarrow \infty $. After we
truncate the higher-order Green's functions, we will encounter related
integrals that by simple manipulation we transform into integrals over
Green's functions $\ll f_{m\sigma },f_{m\sigma }^{+}\gg (\mathcal{G}%
_{m\sigma })$ and $\ll f_{\bar{m}\sigma },f_{m\sigma }^{+}\gg (\mathcal{M}_{%
\bar{m}\sigma })$. Thus the set of EOM terminates after truncation and the
Green's function $\mathcal{G}_{m\sigma }$ can be solved for. The equations
are %\newcounter{saveeqn}\stepcounter{saveeqn}\setcounter{equation}{0}%
\begin{eqnarray}
\sum_{ik,i^{\prime }k^{\prime }}\frac{<c_{i^{\prime }k^{\prime }\bar{m}%
\sigma }^{+}c_{ikm\sigma }>}{z-\varepsilon _{k}} &=&-\frac{4}{4\pi D^{2}}%
\int_{-D}^{D}d\varepsilon \int_{-D}^{D}d\varepsilon ^{\prime }\int
dw^{\prime }f_{FD}(w^{\prime })\frac{\mathrm{Im}\ll c_{ikm\sigma
},c_{i^{\prime }k^{\prime }\bar{m}\sigma }^{+}\gg _{w^{\prime }+i\delta }}{%
z-\varepsilon }  \nonumber \\
&=&\frac{-i\pi }{D^{2}}(V_{0}^{2}\tilde{B}_{m\sigma ,\bar{m}\sigma
}(z)+V_{X}^{2}\tilde{B}_{\bar{m}\sigma ,m\sigma }(z)+V_{0}V_{X}(\tilde{B}%
_{m\sigma ,m\sigma }(z)+\tilde{B}_{\bar{m}\sigma ,\bar{m}\sigma }(z))).
\nonumber \\
\sum_{ik,i^{\prime }k^{\prime }}\frac{<c_{i^{\prime }k^{\prime }m\sigma
}^{+}c_{ikm\sigma }>}{z-\varepsilon _{k}} &=&-\frac{2}{2D}%
\int_{-D}^{D}dw^{\prime }\frac{f_{FD}(w^{\prime })}{w^{\prime }-z-i\delta }+%
\frac{-i\pi }{D^{2}}(V_{0}^{2}\tilde{B}_{m\sigma ,m\sigma }(z)+V_{X}^{2}%
\tilde{B}_{\bar{m}\sigma ,\bar{m}\sigma }(z)  \nonumber \\
&&+V_{0}V_{X}(\tilde{B}_{m\sigma ,\bar{m}\sigma }(z)+\tilde{B}_{\bar{m}%
\sigma ,m\sigma }(z))).  \nonumber \\
\sum_{ik}\frac{<f_{m\sigma }^{+}c_{ikm\sigma }>}{z-\varepsilon _{k}} &=&%
\frac{-2}{2D}(V_{0}\tilde{B}_{m\sigma ,m\sigma }(z))+V_{X}\tilde{B}_{\bar{m}%
\sigma ,m\sigma }(z)).  \nonumber \\
\sum_{ik}\frac{<f_{m\sigma }^{+}c_{ik\bar{m}\sigma }>}{z-\varepsilon _{k}}
&=&\frac{-2}{2D}(V_{0}\tilde{B}_{\bar{m}\sigma ,m\sigma }(z))+V_{X}\tilde{B}%
_{m\sigma ,m\sigma }(z)).  \nonumber \\
\sum_{ik}\frac{<c_{ikm\sigma }^{+}f_{m\sigma }>}{z-\varepsilon _{k}} &=&%
\frac{-2}{2D}(V_{0}\tilde{B}_{m\sigma ,m\sigma }(z))+V_{X}\tilde{B}_{m\sigma
,\bar{m}\sigma }(z)).  \nonumber \\
\sum_{ik}\frac{<c_{ik\bar{m}\sigma }^{+}f_{m\sigma }>}{z-\varepsilon _{k}}
&=&\frac{-2}{2D}(V_{0}\tilde{B}_{m\sigma ,\bar{m}\sigma }(z))+V_{X}\tilde{B}%
_{m\sigma ,m\sigma }(z)).  \nonumber
\end{eqnarray}%
with
\[
\tilde{B}_{\alpha ,\beta }(w)=\int_{-D}^{D}dw^{\prime }f_{FD}(w^{\prime })%
\frac{\ll f_{\alpha },f_{\beta }^{+}\gg ^{\ast }}{w^{\prime }-w-i\delta }
\]%
Note that we have considered both the correlations in the conduction bands,
and on the dot. However, conduction electrons do not mix different valley
states except through cross-valley couplings ($V_{X}$) with the dot. Thus
terms like $\sum_{i^{\prime }k^{\prime },ik}\frac{<c_{i^{\prime }k^{\prime }%
\bar{l}}^{+}c_{ikl}>}{w-\varepsilon _{k}}$ are of order $V_{0(X)}^{2}$,
while terms such as $\sum_{ik}\frac{<f_{l}^{+}c_{ik\bar{l}}>}{w-\varepsilon
_{k}}$ are of order $V_{0(X)}$. Eq.~\ref{eq:f4} and \ref{eq:ff4} can be thus
expressed as
\begin{eqnarray}
(w-\varepsilon _{m\sigma }-U-\Sigma _{a}(w)-\Sigma _{a}(z))\ll
n_{l}f_{m\sigma },f_{m\sigma }^{+}\gg =\Sigma _{b}(w)\ll n_{l}f_{%
\bar{m}\sigma },f_{m\sigma }^{+}\gg _{l\not=\bar{m}\sigma } \nonumber
 \\ +\{\delta _{l,\bar{m}\sigma }(\widetilde{I}_{m\sigma }(w)-\Sigma _{a}(w)%
\widetilde{F}_{m\sigma }(w) +\Sigma _{d}(w))+\Sigma _{b}(w)\tilde{A}_{l}(z)\}%
\mathcal{M}_{\bar{m}\sigma } \nonumber \\+\{-\delta _{l,\bar{m}\sigma }\Sigma _{b}(w)%
\widetilde{F}_{m\sigma }(w)
+\Sigma _{a}(w)\tilde{A}_{l}(z)-\Sigma _{c,l}(z)-\widetilde{J}_{l}(z)\}%
\mathcal{G}_{m\sigma }+<n_{l}>+\tilde{A}_{l}(z)  \label{eq:rf4} \\
(w-\varepsilon _{\bar{m}\sigma }-U-\Sigma _{a}(w)-\Sigma _{a}(\bar{z}))\ll
n_{p}f_{\bar{m}\sigma },f_{m\sigma }^{+}\gg =\Sigma _{b}(w)\ll
n_{p}f_{m\sigma },f_{m\sigma }^{+}\gg _{p\not=m\sigma }  \nonumber \\
-\delta _{p,m\sigma }<f_{m\sigma }^{+}f_{\bar{m}\sigma }>-\delta
_{p,m\sigma }\widetilde{\bar{F}}_{m\sigma }(w)+\{\delta _{p,m\sigma }(%
\widetilde{\bar{I}}_{m\sigma }(w)-\Sigma _{a}(w)\widetilde{\bar{F}}_{m\sigma
}(w)+\Sigma _{d}(w))  \nonumber \\
+\Sigma _{b}(w)\tilde{A}_{p}(\bar{z})\}\mathcal{G}_{m\sigma }+\{-\delta
_{p,m\sigma }\Sigma _{b}(w)\widetilde{\bar{F}}_{m\sigma }(w)+\Sigma _{a}(w)%
\tilde{A}_{p}(\bar{z})-\Sigma _{c,p}(\bar{z})-\widetilde{J}_{p}(\bar{z})\}%
\mathcal{M}_{\bar{m}\sigma }  \label{eq:rff4}
\end{eqnarray}%
where the various functions are defined by
\begin{eqnarray}
\Sigma _{a}(w) &=&\sum_{ik}\frac{V_{0}^{2}+V_{X}^{2}}{w-\varepsilon _{k}}
\label{eq:deffunction_1} \\
\Sigma _{b}(w) &=&\sum_{ik}\frac{2V_{0}V_{X}}{w-\varepsilon _{k}} \\
\Sigma _{c,l}(z) &=&\sum_{i^{\prime }k^{\prime }ik}\frac{%
(V_{0}^{2}+V_{X}^{2})<c_{i^{\prime }k^{\prime }l}^{+}c_{ikl}>}{z-\varepsilon
_{k}}  \nonumber \\
&=&-\frac{2}{2D}\int_{-D}^{D}dw^{\prime }f_{FD}(w^{\prime })\frac{%
V_{0}^{2}+V_{X}^{2}}{w^{\prime }-z-i\delta }+\frac{-\pi
(V_{0}^{2}+V_{X}^{2})i}{D^{2}}(V_{0}^{2}\tilde{B}_{l,l}(z)+V_{X}^{2}\tilde{B}%
_{\bar{l},\bar{l}}(z)  \nonumber \\
&&+V_{0}V_{X}(\tilde{B}_{l,\bar{l}}(z)+\tilde{B}_{\bar{l},l}(z))). \\
\Sigma _{d}(w) &=&\sum_{ik,i^{\prime }k^{\prime }}\frac{2V_{0}V_{X}<c_{i^{%
\prime }k^{\prime }m\sigma }^{+}c_{ikm\sigma }>}{w-\varepsilon _{k}}
\nonumber \\
&=&-\frac{2}{2D}\int_{-D}^{D}dw^{\prime }f_{FD}(w^{\prime })\frac{2V_{0}V_{X}%
}{w^{\prime }-w-i\delta }+\frac{-2\pi V_{0}V_{X}i}{D^{2}}(V_{0}^{2}\tilde{B}%
_{m\sigma ,m\sigma }(w)+V_{X}^{2}\tilde{B}_{\bar{m}\sigma ,\bar{m}\sigma }(w)
\nonumber \\
&&+V_{0}V_{X}(\tilde{B}_{m\sigma ,\bar{m}\sigma }(w)+\tilde{B}_{\bar{m}%
\sigma ,m\sigma }(w))).  \label{fn:sma_d} \\
\widetilde{I}_{m\sigma }(w) &=&\sum_{i^{\prime }k^{\prime }i k}\frac{%
V_{0}^{2}<c_{i^{\prime }k^{\prime }\bar{m}\sigma }^{+}c_{ikm\sigma
}>+V_{X}^{2}<c_{i^{\prime }k^{\prime }m\sigma }^{+}c_{ik\bar{m}\sigma }>}{%
w-\varepsilon _{k}}  \nonumber \\
&=&\frac{-i\pi }{D^{2}}((V_{0}^{4}+V_{X}^{4})\tilde{B}_{m\sigma ,\bar{m}%
\sigma }(w)+2V_{0}^{2}V_{X}^{2}\tilde{B}_{\bar{m}\sigma ,m\sigma }(w)
\nonumber \\
&&+(V_{0}^{2}+V_{X}^{2})V_{0}V_{X}(\tilde{B}_{m\sigma ,m\sigma }(w)+\tilde{B}%
_{\bar{m}\sigma ,\bar{m}\sigma }(w))). \\
\widetilde{F}_{m\sigma }(w) &=&\sum_{ik}\frac{V_{0}<f_{\bar{m}\sigma
}^{+}c_{ikm\sigma }>+V_{X}<f_{\bar{m}\sigma }^{+}c_{ik\bar{m}\sigma }>}{%
w-\varepsilon _{k}}  \nonumber \\
&=&\frac{-2}{2D}((V_{0}^{2}+V_{X}^{2})\tilde{B}_{m\sigma ,\bar{m}\sigma
}(w)+2V_{0}V_{X}\tilde{B}_{\bar{m}\sigma ,\bar{m}\sigma }(w)). \\
\widetilde{J}_{l}(z) &=&\sum_{i^{\prime }k^{\prime }i k}\frac{%
V_{0}V_{X}(<c_{i^{\prime }k^{\prime }\bar{l}}^{+}c_{ikl}>+<c_{i^{\prime
}k^{\prime }l}^{+}c_{ik\bar{l}}>)}{z-\varepsilon _{k}}  \nonumber \\
&=&\frac{-i\pi }{D^{2}}V_{0}V_{X}((V_{0}^{2}+V_{X}^{2})(\tilde{B}_{\bar{l}%
,l}(z)+\tilde{B}_{l,\bar{l}}(z))+2V_{0}V_{X}(\tilde{B}_{l,l}(z)+\tilde{B}_{%
\bar{l},\bar{l}}(z))). \\
\tilde{A}_{l}(z) &=&\sum_{ik}\frac{V_{X}<f_{l}^{+}c_{ik\bar{l}%
}>+V_{0}<f_{l}^{+}c_{ikl}>}{z-\varepsilon _{k}}  \nonumber \\
&=&\frac{-2}{2D}((V_{0}^{2}+V_{X}^{2})\tilde{B}_{l,l}(z)+2V_{0}V_{X}\tilde{B}%
_{\bar{l},l}(z)). \\
\widetilde{\bar{I}}_{m\sigma }(w) &=&\sum_{i^{\prime }k^{\prime }i k}\frac{%
V_{0}^{2}<c_{i^{\prime }k^{\prime }m\sigma }^{+}c_{ik\bar{m}\sigma
}>+V_{X}^{2}<c_{i^{\prime }k^{\prime }\bar{m}\sigma }^{+}c_{ikm\sigma }>}{%
w-\varepsilon _{k}}  \nonumber \\
&=&\frac{-i\pi }{D^{2}}((V_{0}^{4}+V_{X}^{4})\tilde{B}_{\bar{m}\sigma
,m\sigma }(w)+2V_{0}^{2}V_{X}^{2}\tilde{B}_{m\sigma ,\bar{m}\sigma }(w)
\nonumber \\
&&+(V_{0}^{2}+V_{X}^{2})V_{0}V_{X}(\tilde{B}_{m\sigma ,m\sigma }(w)+\tilde{B}%
_{\bar{m}\sigma ,\bar{m}\sigma }(w))). \\
\widetilde{\bar{F}}_{m\sigma }(w) &=&\sum_{ik}\frac{V_{0}<f_{m\sigma
}^{+}c_{ik\bar{m}\sigma }>+V_{X}<f_{m\sigma }^{+}c_{ikm\sigma }>}{%
w-\varepsilon _{k}}  \nonumber \\
&=&\frac{-2}{2D}((V_{0}^{2}+V_{X}^{2})\tilde{B}_{\bar{m}\sigma ,m\sigma
}(w)+2V_{0}V_{X}\tilde{B}_{m\sigma ,m\sigma }(w)).  \label{eq:deffunction_10}
\end{eqnarray}%
Combining Eq.~\ref{eq:aa1},~\ref{eq:aa2},~\ref{eq:rf4}, \ref{eq:rff4}, and
taking the limit $U\rightarrow \infty $, we obtain Eq.~\ref{eq:fbarf} in the
main text.


\begin{thebibliography}{99}
\bibitem{Jantsch2002} W. Jantsch \textit{et al}, Physica E \textbf{13}, 504
(2002); Z. Wilamowski \textit{et al} Physica E \textbf{16}, 111 (2003).

\bibitem{Tyryshkin2005} A. M. Tyryshkin, S. A. Lyon, W. Jantsch, F.
Schaeffler, Phys. Rev. Lett. \textbf{94}, 126802 (2005).

\bibitem{Tyryshkin2003} A. M. Tyryshkin, S. A. Lyon, W. Jantsch, F.
Schaeffler, Phys. Rev. B \textbf{68}, 193207 (2003).

\bibitem{DasSarma2003} R. de Sousa and S. Das Sarma, Phys. Rev. B \textbf{68}%
, 115322 (2003)

\bibitem{Eriksson2004} M. A. Eriksson \textit{et al.} Quantum Information
Processing \textbf{3}, 133 (2004).

\bibitem{Jones2006} G. M. Jones \textit{et al.} Appl. Phys. Lett. \textbf{89}%
, 073106 (2006).

\bibitem{Klein2004} L. J. Klein \textit{et al.} Appl. Phys. Lett. \textbf{84}%
, 4047 (2004).

\bibitem{srijit_nature} S. Goswami \textit{et al.}, cond-mat/0611221. To
appear in Nature Physics.

\bibitem{TBBoykin2004_2} T. B. Boykin \textit{et al.} Phys. Rev. B \textbf{70%
}, 165325 (2004).

\bibitem{TBBoykin2004} T. B. Boykin \textit{et al.} Appl. Phys. Lett.
\textbf{84}, 115 (2004).

\bibitem{Ohkawa1977} F. J. Ohkawa and Y. Uemura, Journ. Phys. Soc. Japan,
\textbf{43}, 925 (1977).

\bibitem{friesen1} M. Friesen, S. Chutia, C. Tahan, and S.N. Coppersmith,
cond-mat/0608229.

\bibitem{Friesen2006} Mark Friesen, M. A. Eriksson, and S. N. Coppersmith,
cond-mat/0602194.

\bibitem{Goldhaber1998} D. Goldhaber-Gordon \textit{et al.} Nature \textbf{%
391}, 156 (1998).

\bibitem{Jarillo} P. Jarillo-Herrero \textit{et al.} Nature(London) \textbf{%
434}, 484 (2005).

\bibitem{Choi2005} M.-S. Choi \textit{et al.} Phys. Rev. Lett. \textbf{95},
067204 (2005).

\bibitem{Sasaki} S. Sasaki \textit{et al.} Phys. Rev. Lett. \textbf{93},
017205 (2004).

\bibitem{tsui} L. P. Rokhinson, L. J. Guo, S. Y. Chou, and D. C. Tsui, Phys.
Rev B \textbf{60,} R16319 (1999)


\bibitem{JSLim2006} J. S. Lim \textit{et al.} cond-mat/0608110.

\bibitem{orbitkondo} T. Inoshita et al., Phys. Rev. B \textbf{48}, 14725
(1993); T. Pohjola \textit{et al}., Europhys. Lett. \textbf{40}, 189 (1997);
A. Levy Yeyati \textit{et al}., Phys. Rev. Lett. \textbf{83}, 600 (1999); R.
Sakano and N. Kawakami, Phys. Rev. B \textbf{73}, 155332 (2006).


\bibitem{Chudnovskiy2005} A. L. Chudnovskiy, Europhys. Lett. \textbf{71},
672 (2005).


\bibitem{Landauer} R. Landauer, Philos. Mag. \textbf{21}, 863 (1970).

\bibitem{Czycholl} G. Czycholl, Phys. Rev. B \textbf{31}, 2867 (1985).


\bibitem{Lacroix1981} C. Lacroix, J. Phys. F. \textbf{11}, 2389 (1981).

\bibitem{Luo2002} H.-G. Luo, S.-J. Wang, and C.-L. Jia, Phys. Rev. B \textbf{%
66}, 235311 (2002).

\bibitem{Kashcheyevs2006} V. Kashcheyevs, A. Aharony, and O. Entin-Wohlman,
Phys. Rev. B \textbf{73}, 125338 (2006).

%\bibitem{Meir1991} Y. Meir, N. Wingreen, and P. Lee, Phys. Rev. Letts {\bf 66}, 3408 (1991).
%\bibitem{mahan} Gerald D. Mahan, {\it Many-Particle Physics, Third Edition}.

\bibitem{foot_note1} In the case where $\beta>0$, mathematically there is no
constraint to the magnitude of $<n_{m\sigma}>$ and $<f^+_{m\sigma}f_{\bar{m}%
\sigma}>$ when iterating the computation to determine their values. However,
note that the value $<f^+_{m\sigma}f_{\bar{m}\sigma}>$ should be
approximately of order $O(\beta\Gamma<n_{m\sigma}>)$. Since the stability
doesn't reach at $<f^+_{m\sigma}f_{\bar{m}\sigma}>\sim
O(\beta\Gamma<n_{m\sigma}>)$, we first stablize $<n_{m\sigma}>$ by setting $%
<f^+_{m\sigma}f_{\bar{m}\sigma}>=0$ in a sense that we treat $%
<f^+_{m\sigma}f_{\bar{m}\sigma}>$ perturbatively. Afterwards we iterate to
determine $<f^+_{m\sigma}f_{\bar{m}\sigma}>$ and stop when it reaches a
value $\sim \beta \Gamma<n_{m\sigma}>$ and take both values to plot the DOS.

\bibitem{Hewson} A.P. Hewson, \textit{The Kondo Problem to Heavy Fermions, }%
(Cambridge Univ. Press, Cambridge, 1993), Sec. 7.2.

\bibitem{Eto2005} M. Eto, J. Phys. Soc. Jpn. \textbf{74}, 95 (2005).

%\bibitem{srijit2006} Srijit Goswami \textit{et al.} cond-mat/0408389.
\end{thebibliography}
\end{document}